\documentclass[twocolumn,prb,showpacs,floatfix]{revtex4}
\usepackage{graphicx}
\usepackage{bm}

\begin{document}

\title{\textbf{Spin dynamics of frustrated easy-axis triangular
antiferromagnet $2H$-AgNiO$_2$
explored~by~inelastic~neutron~scattering}}
\author{E.~M.~Wheeler$^{1,2}$}
\altaffiliation[Present address: ]{Helmholtz Centre Berlin for
Materials and Energy, Glienicker Str. 100, D-14109 Berlin,
Germany}
\author{R.~Coldea$^{3}$}
\author{E.~Wawrzy\'{n}ska$^{3}$}
\author{T.~S\"{o}rgel$^{4}$}
\author{M.~Jansen$^{4}$}
\author{M.~M.~Koza$^{2}$}
\author{J.~Taylor$^{5}$}
\author{P.~Adroguer$^{3,6}$}
\author{N.~Shannon$^{3}$}
\affiliation{$^1$Clarendon Laboratory, University of Oxford, Parks
Road, Oxford OX1 3PU, United Kingdom\\
$^2$Institut Laue-Langevin, BP 156, 38042 Grenoble Cedex 9, France\\
$^3$H.H. Wills Physics Laboratory, University of Bristol, Tyndall
Avenue, Bristol, BS8 1TL, United Kingdom \\
$^4$Max-Planck Institut f\"{u}r Festk\"{o}rperforschung,
 Heisenbergstrasse 1, D-70569 Stuttgart, Germany \\
$^5$ISIS Facility, Rutherford Appleton Laboratory, Chilton, Didcot
OX11 0QX, United Kingdom\\
$^6$Laboratoire de Physique, \'{E}cole normale sup\'{e}rieure de
Lyon, 46 All\'{e}e d'Italie, 69364 Lyon Cedex 07, France}
\date{\today}
\pacs{75.25.+z, 75.10.Jm}

\begin{abstract}
We report inelastic neutron scattering measurements of the spin
dynamics in the layered hexagonal magnet $2H$-AgNiO$_2$, which has
stacked triangular layers of antiferromagnetically-coupled
Ni$^{2+}$ spins ($S$=1) ordered in a collinear alternating stripe
pattern. We observe a broad band of magnetic excitations above a
small gap of 1.8\ meV and extending up to 7.5\ meV, indicating
strongly dispersive excitations. The measured dispersions of the
boundaries of the powder-averaged spectrum can be quantitatively
explained by a linear spin-wave dispersion for triangular layers
with antiferromagnetic nearest- and weak next-nearest neighbor
couplings, a strong easy-axis anisotropy and additional weak
inter-layer couplings. The resulting dispersion relation has
global minima not at magnetic Bragg wavevectors but at
symmetry-related soft points and we attribute this anomalous
feature to the strong competition between the easy-axis anisotropy
and the frustrated antiferromagnetic couplings. We have also
calculated the quantum corrections to the dispersion relation to
order $1/S$ in spin-wave theory by extending the work of Chubukov
and Jolicoeur [Phys. Rev. B \textbf{46}, 11137 (1992)] and find
that the presence of easy-axis anisotropy significantly reduces
the quantum renormalizations predicted for the isotropic model.
\end{abstract}

\maketitle

\section{Introduction}
\label{sec_introduction}
\begin{figure}[tb]
\begin{center}
\includegraphics[width=7cm,bbllx=203,bblly=451,bburx=428,
  bbury=680,angle=0,clip=]{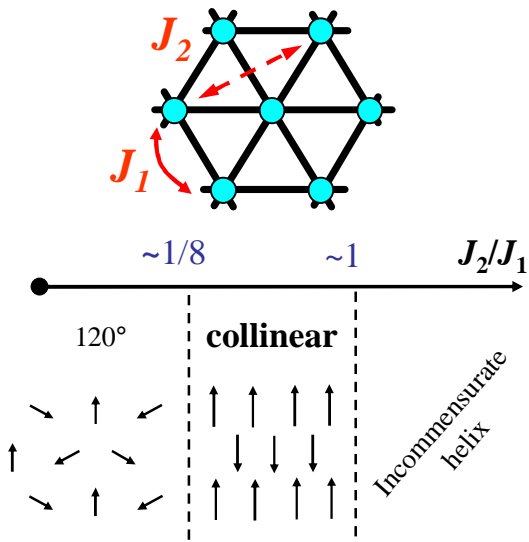}
  \caption{\label{fig_J1J2} (Color online) Phases of the Heisenberg triangular
  lattice antiferromagnet ($120^{\circ}$ coplanar order, collinear
  stripes and incommensurate spiral) as a function of the ratio between the
  first and second neighbor couplings, $J_1$ and $J_2$
  respectively. In the collinear phase the common spin direction
  is spontaneously chosen and can point anywhere, it is shown here
  in-plane for ease of visualization.}
\end{center}
\end{figure}

Quantum antiferromagnets on triangular lattices provide model
systems for investigating the effects of quantum fluctuations and
geometric frustration. Zero point fluctuations are expected to be
strongly enhanced for low spin and frustrated lattice geometry and
may stabilize non-classical ordered or spin liquid phases, or
unconventional spin dynamics. The triangular Heisenberg
antiferromagnet with first and second neighbor couplings $J_1$ and
$J_2$ [see Fig.~\ref{fig_J1J2}] shows strong frustration effects
and a macroscopically-degenerate classical ground state for $1/8
\le J_2/J_1 \le 1$. This was initially proposed theoretically as a
candidate for a chiral spin liquid state.\cite{Baskaran_1989}
However, perturbative expansions using a spin-wave
basis\cite{Jolicoeur_1990,Chubukov_1992,Lecheminant_1992}
predicted that in this range of couplings quantum fluctuations do
not stabilize a spin liquid state but instead lift the classical
degeneracy to select a collinear stripe order pattern [see Fig~\
\ref{fig_J1J2}]. Moreover, large quantum renormalizations of the
spin-wave dispersion relations compared to classical have been
predicted, but not probed experimentally in the absence of a good
experimental model system. The recent observation of a collinear
stripe-ordered ground state in the layered hexagonal
antiferromagnet $2H$-AgNiO$_2$ [ref.~\onlinecite{AgNiO2_prl}]
suggests a potential realization of the experimentally-unexplored
frustrated $J_1-J_2$ model in the collinear phase and here we
present first inelastic neutron scattering measurements which give
information about the spin gap and dispersion relations. We find
that the data can be parameterized by a spin-wave dispersion
relation for a $J_1$-$J_2$ triangular antiferromagnet with
strong easy-axis anisotropy and weak inter-layer couplings.

\begin{figure}[tb]
\begin{center}
  \includegraphics[width=8cm,bbllx=55,bblly=265,bburx=540,
  bbury=575,angle=0,clip=]{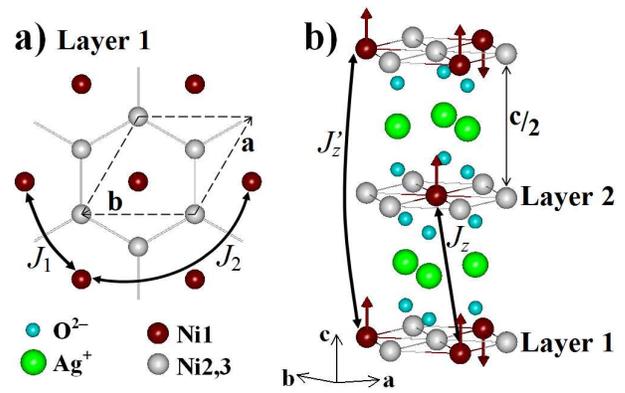}
  \caption{\label{fig_Structure}(Color online) a) Triangular lattice
  formed by Ni1 sites (dark brown spheres) in a hexagonal NiO$_2$ layer
  in $2H$-AgNiO$_2$; thick arrows indicate the paths for the
  in-plane nn and nnn exchanges $J_1$ and $J_2$. Dashed diamond shows the
  crystallographic unit cell. Gray spheres are Ni2 and Ni3
  sites, assumed non-magnetic.
  b) 3D view of the crystal and magnetic structure. There are two
  symmetry-equivalent NiO$_2$ layers per unit cell. Layer stacking
  is such that Ni1 ions sit above the centre of a Ni1 triangle in
  the layer below. We consider two natural candidates for inter-layer
  couplings: $J_z$ between Ni1 ions in adjacent layers (three neighbours
  above and three below) and $J_z'$ between Ni1 ions at two layers apart.}
\end{center}
\end{figure}

Delafossite materials of the type $X$NiO$_2$ have been generally
thought of as possible two-dimensional frustrated magnets. A
network of edge-sharing NiO$_6$ octahedra lead to a triangular
lattice arrangement of Ni ions in planes spaced by layers of $X^+$
ions. There have been a variety of studies investigating the
properties of compounds where $X$=Li, Na. LiNiO$_2$ has shown no
long-range magnetic order but experiments are hindered by the
difficulty in preparing stoichiometric samples.\cite{Chung_2005}
From elastic and inelastic neutron scattering measurements,
NaNiO$_2$ was found to be a spin-1/2 system with in-plane
(un-frustrated) ferromagnetic interactions and weak
antiferromagnetic inter-layer couplings.\cite{Lewis_2005}

We have recently started exploring the delafossite AgNiO$_2$,
which shows frustrated antiferromagnetic in-plane
interactions\cite{agnio2early,soergel}. Detailed structural
studies have been performed on the hexagonal polytype of
AgNiO$_2$, so called $2H$-AgNiO$_2$ with two NiO$_2$ layers per
unit cell [see Fig.~\ref{fig_Structure}b)], as opposed to the
earlier-synthesized rhombohedral $3R$ polytype\cite{agnio2early}
with a 3-layer stacking sequence along $c$-axis. High-resolution
neutron diffraction measurements of $2H$-AgNiO$_2$ have observed a
structural transition upon cooling below 365 K, which leads to a
periodic arrangement of expanded and contracted NiO$_6$
octahedra.\cite{AgNiO2_mag_order} This was
proposed\cite{AgNiO2_prl} to be a consequence of spontaneous
charge order on the Ni sites driven by the need to lift a two-fold
orbital degeneracy. This leads to a strongly-magnetic Ni1 site
(spin-1 Ni$^{2+}$, 1/3$^{\rm rd}$ of sites) arranged in an ideal
triangular lattice [see Fig.~\ref{fig_Structure}a)], with the
remaining 2/3$^{\rm rds}$ of Ni sites being Ni$^{3.5+}$ with an
itinerant character.

\begin{figure}[tb]
\begin{center}
  \includegraphics[width=8cm,bbllx=54,bblly=335,bburx=539,
    bbury=505] {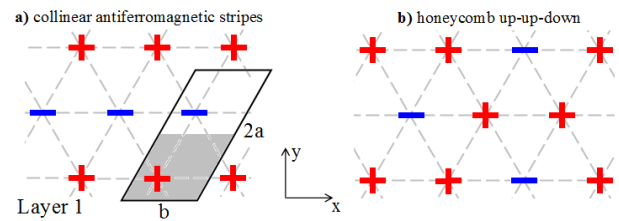}
  \caption{\label{fig_Agnio2_mag_structure}(Color online) a) Magnetic
  structure in a triangular layer showing the alternating stripe order
  of the Ni1 spins ($\pm$ symbols indicate the projection of the ordered
  spin moments along the $c$-axis). $x$, $y$ labels indicate a
  natural orthogonal coordinate system with $x$ along and $y$ transverse
  to stripes. Solid diamond indicates the magnetic unit cell, doubled
  along the crystallographic $a$-axis compared to the chemical unit
  cell (light gray shaded area). The depicted magnetic structure has
  stripes running along the $b$-axis and ordering wavevector $\bm{k}=(1/2,0,0)$;
  equivalent domains are obtained by rotating this pattern by $\pm60^{\circ}$
  around the $c$-axis. Further neighbor couplings are required to stabilize
  this structure with respect to the ferrimagnetic honeycomb up-up-down
  pattern shown in b).}
\end{center}
\end{figure}

Magnetic order occurs below $T_N=19.7(3)$ K\cite{AgNiO2_mag_order}
when the magnetic moments of Ni1 sites order in a collinear
pattern of alternating ferromagnetic rows (stripes) in the
triangular layers with spins pointing along the $c$-axis, as shown
in Fig.~\ref{fig_Agnio2_mag_structure}a). A large ordered moment
is found on the Ni1 sites, 1.552(7) $\mu_{\rm B}$ at 4 K, and no
significant ordered moment could be detected on the remaining Ni2
and Ni3 sites, proposed by band-structure
calculations\cite{AgNiO2_prl} to be strongly itinerant and
possibly ordered but with only a very small
($\rlap{\lower4pt\hbox{\hskip1pt$\sim$}}
    \raise1pt\hbox{$<$}~ 0.1 \mu_{\rm B}$) moment. So to a
first approximation the coherent spin dynamics in the
magnetically-ordered phase is expected to be dominated by the
$S=1$ Ni1 spins arranged on an ideal triangular lattice with
antiferromagnetic couplings.

The experimentally observed collinear stripe order in
Fig.~\ref{fig_Agnio2_mag_structure}a) is rather unusual for a
triangular antiferromagnet (TAFM), but has been proposed to occur
for the classical Ising model for finite second-neighbor couplings
$J_2 \ge 0$ [Ref. \onlinecite{Slotte_1984}]. For the Heisenberg
TAFM the collinear stripe order has also been proposed to occur as
a ground state for moderate $J_2$ in the range $1/8
~\rlap{\lower4pt\hbox{\hskip1pt$\sim$}}
    \raise1pt\hbox{$<$}~J_2/J_1
~\rlap{\lower4pt\hbox{\hskip1pt$\sim$}}
    \raise1pt\hbox{$<$}~1$, Ref. \onlinecite{Chubukov_1992}.
    In this range the classical ground state is
macroscopically degenerate and there are many other non-collinear
states degenerate with the two-sublattice collinear stripe order
shown in Fig.~\ref{fig_Agnio2_mag_structure}a), but zero-point
quantum fluctuations are predicted to select the latter through
the ``order by disorder" mechanism. The large classical degeneracy
is manifested also in the linear spin-wave dispersion which has
many soft points with zero energy, and proper inclusion of quantum
corrections leads to renormalizations of the semiclassical
dispersion relation and a gapping of the non-physical zero
modes.\cite{Chubukov_1992} Those effects have remained
experimentally largely unexplored. The recent
observation\cite{AgNiO2_prl} of a collinear stripe ordered phase
in the hexagonal magnet AgNiO$_2$ suggested a possible
experimental realization of the $S=1$ TAFM in the range of
moderate frustration and motivated us to measure its spin dynamics
in some detail. We note that the physics we are exploring here
appears to be different to that in the recently explored $S=5/2$
TAFM CuFeO$_2$ which has a related, but different collinear
structure of alternating ``double stripes"
($\uparrow\uparrow\downarrow\downarrow$) stabilized by strong
three-dimensional and third-neighbour in-plane
couplings.\cite{CuFeO2} In $2H$-AgNiO$_2$ the magnetic structure
is a (simpler) single-stripe pattern $\uparrow\downarrow$ with
only two in-plane sublattices possibly stabilized by fewer
exchanges (only first and second in-plane neighbours) and
easy-axis anisotropy.

The plan of this paper is as follows. The following Sec.\
\ref{sec_exp_details} gives details of the inelastic neutron
scattering experiments to probe the powder-averaged spin dynamics
and the results are presented in Sec.\ \ref{sec_results}. In the
following Sec.\ \ref{sec_sinusoidal} the data is parameterized in
terms of an empirical sinusoidal dispersion model with minima at
magnetic Bragg wavevectors and different zone boundary energies
along the three orthogonal directions in the Brillouin zone; this
parameterization gives a gapped and predominantly two-dimensional
dispersion relation. In the following Sec.\
\ref{sec_coupled_layers} the data is compared with linear
spin-wave theory for a microscopic spin Hamiltonian for the
localized Ni spins that includes both first- and second-neighbour
antiferromagnetic couplings in the triangular layers, easy-axis
anisotropy modelled by a single-ion term, and different models for
inter-layer couplings consistent with the crystal structure; a
minimal spin Hamiltonian is proposed. In the following Sec.\
\ref{sec_swt} we consider how magnon interactions included at
order $1/S$ in spin-wave theory renormalize the dispersion
relation and find slightly renormalized values for the proposed
spin interactions. Finally, the results for the spin dynamics are
summarized and discussed in the concluding Sec.\
\ref{sec_Conclusions}. Technical details of the spin wave
calculations for coupled easy-axis triangular layers are given in
Appendix\ \ref{sec_appendix}. Details of the derivation of the
$1/S$ quantum corrections to the dispersion relation for an
easy-axis triangular antiferromagnet are given in Appendix\
\ref{sec_appendixB}.

\section{Experimental Details}
\label{sec_exp_details}

Powder samples of the hexagonal polytype of AgNiO$_{2}$ were
prepared from Ag$_{2}$O and Ni(OH)$_{2}$ using high oxygen
pressures (130 MPa) as described in Ref. \onlinecite{soergel}. The
samples used for the inelastic neutron experiments were part of
the same batch used in previous neutron diffraction
measurements\cite{AgNiO2_mag_order} performed on about half of the
total sample quantity, which revealed a high-purity hexagonal
phase ($< 1\%$ admixture of the rhombohedral polytype). The
crystal structure of $2H$-AgNiO$_2$ is shown in
Fig.~\ref{fig_Structure}b) and is hexagonal with space group P$6_3
2 2$ (no. 182) and lattice parameters $a=5.0908(1)$\AA\ and
$c=12.2498(1)$\AA\ [Ref. \onlinecite{AgNiO2_prl}]. Ni ions are
located inside hexagonal ($ab$) layers and there are three
distinct crystallographic sites: Ni1 which sit inside slightly
expanded NiO$_6$ octahedra arranged in a periodic triangular
lattice of spacing $a$, surrounded by a honeycombe of contracted
NiO$_6$ octahedra which contain the Ni2 and Ni3 sites, see
Fig.~\ref{fig_Structure}a).

The powder-averaged magnetic excitation spectrum was probed using
two direct-geometry time-of-flight neutron spectrometers: MARI, at
the ISIS Facility in the UK, and IN6, at the Institute
Laue-Langevin in France. Measurements on MARI showed that the full
dynamic range of the spin excitations extended only up to 7.5\ meV
so could be accessed with incident neutrons of energy $E_i=18$\
meV, which gave an energy resolution of 0.61(1) meV (FWHM) on the
elastic line. Higher-resolution measurements to probe the
low-energy part of the spectrum were made using IN6 operated with
incident neutrons of energy $E_i=3.86$ meV, which gave a measured
energy resolution of 0.142(1) meV (FWHM) on the elastic line. The
sample was cooled using either a closed-cycle refrigerator (MARI,
base temperature 4.7\ K) or orange cryostat (IN6, base temperature
1.8\ K). Measurements were made at the lowest temperatures in the
magnetically ordered phase, near the ordering transition and at
high temperatures in the paramagnetic phase.

For the MARI experiment 23 g of powder were placed in an
annular-shaped sachet to minimize absorption. For the low-energy
IN6 measurements a similar powder quantity was placed in a
plate-shaped container angled at $45^\circ$ to the incident beam
and the data was corrected for neutron absorption effects using a
numerical calculation for plate-shaped samples. The scattering
intensities from both instruments have been converted into
absolute units of $S(Q,\omega)$ of mbarns/meV/sr/Ni1 by
normalizing the raw counts to the sample mass and to the measured
scattering intensities from a Vanadium standard.

\begin{figure*}[ht!]
\begin{center}
\includegraphics*[width=17.5cm]{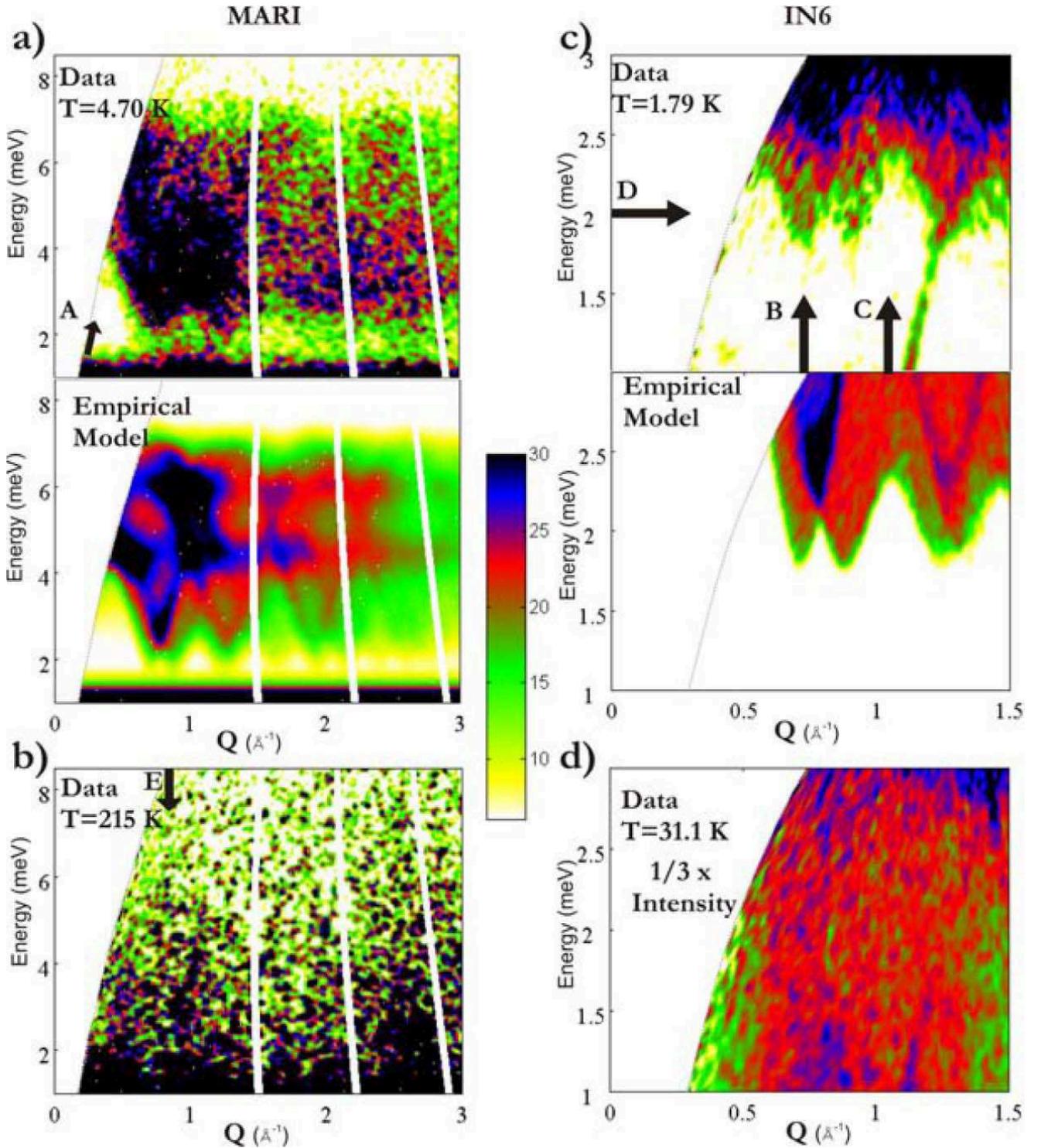}

  \caption{\label{fig_IN6_MARI_mslice}(Color online)
Powder-averaged magnetic excitation spectrum in $2H$-AgNiO$_2$ in
the magnetically-ordered phase (top panels a) and c)) showing a
band of spin excitations above a small gap, b) in the paramagnetic
phase at high temperature showing an overdamped signal and d)
slightly above the magnetic ordering temperature showing a
filling-in of the spin gap [compare with c)]. Data is raw counts
collected after 35 hours (a) and 4.5 hours (c) and normalized to
absolute units as described in the text. The streak of intensity
in panel c) near $Q=1.1$\ \AA$^{-1}$ is non-magnetic and
attributed to an acoustic phonon dispersing out of the (002)
structural Bragg peak. Middle panels show the intensity
distribution for the empirical dispersion model in eqs.\
(\ref{Eq:dispersion}) and (\ref{Eq:E1E2E2_Sqw}). The calculations
include convolution with the instrumental resolution, the magnetic
form factor, polarization factor and an estimate of the
non-magnetic background to be directly compared with the data in
the panels immediately above. Thick bold arrows labelled A-E
indicate the location of scans plotted in Fig.
\ref{fig_E1E2E3_Simulation_results}A-E and the grey dotted lines
in all plots indicate the low-$Q$ edge of the measured region.}
  \end{center}
\end{figure*}

\section{Measurements and Results}
\label{sec_results}

\begin{figure*}[tbh]
\begin{center}
\includegraphics[width=16cm] {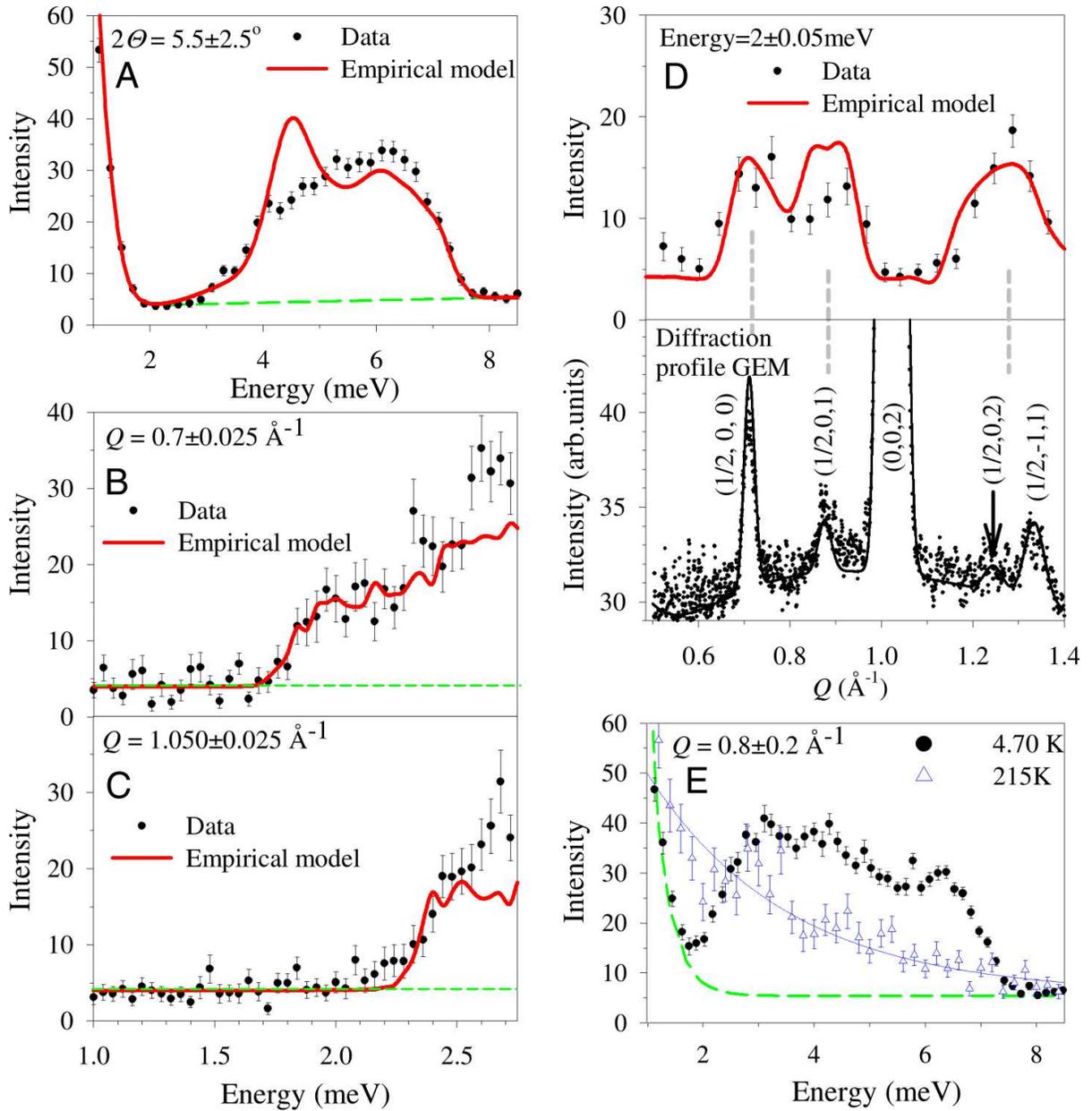} \caption{ \label{fig_E1E2E3_Simulation_results}(Color
online) Scans through the powder data along special directions
labelled A-E in Fig.~\ref{fig_IN6_MARI_mslice}. Scan D shows that
there is structure in the low-energy inelastic signal for
wavevectors where Bragg peaks occur (lower panel in D shows
diffraction data indicating magnetic Bragg peaks). The energy scan E
integrates over a broad $Q$-range and shows the full bandwidth of
the magnetic excitations: low temperature data (solid points) is
replaced in the paramagnetic phase at high temperatures by an
overdamped signal (open triangles, solid line is guide to the eye).
In panels A-D the thick solid lines show calculations for the
sinusoidal dispersion model in eqs.\ (\ref{Eq:dispersion}) and
(\ref{Eq:E1E2E2_Sqw}) with dispersion relations plotted in Fig.\
\ref{fig_Panneled_Dispersion}. The green dashed lines indicate the
estimated non-magnetic background.}
  \end{center}
\end{figure*}

\begin{figure}[tbh]
\begin{center}
\includegraphics[width=7cm,  bbllx=176,bblly=331,bburx=423,  bbury=520]
 {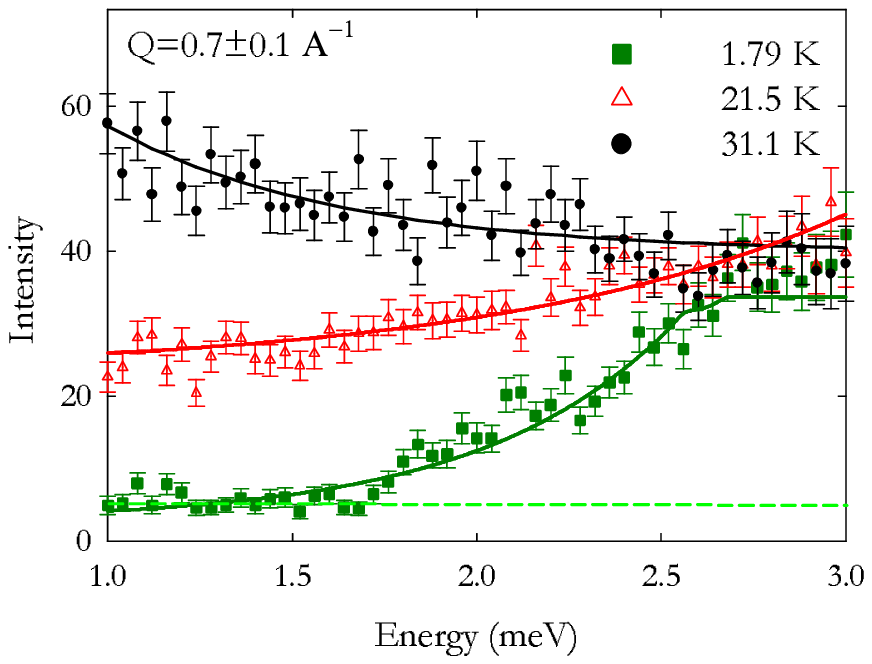}
  \caption{\label{fig_Energy_gap_IN6_with_Temp} (Color online)
  Energy scan at the minimum gap wavevector showing a filling
  of the gap upon heating above the magnetic ordering temperature.
  Solid lines are guides to the eye and the dashed line indicates
  the estimated non-magnetic background.}
  \end{center}
\end{figure}

An overview of the measured inelastic neutron scattering data is
shown in Fig.~\ref{fig_IN6_MARI_mslice}. Below the N\'{e}el
ordering temperature $T_N$=19.7(3)K a strong band of scattering is
clearly observed at low energies, $E$, and low wavevectors,
$Q=|\bm{Q}|$, [Fig.~\ref{fig_IN6_MARI_mslice}a) and c)],
attributed to magnetic excitations. The intensity of the signal
decreases with increasing wavevector $Q$, confirming its magnetic
character. The intensity starts above a small gap of $1.8$ meV and
extends up to $7.5$ meV, above which the signal decreases to
background level. This energy scale for the magnetic excitations
spectrum is consistent with susceptibility measurements which
observed a Curie-Weiss temperature of $k_{\rm B}\theta=-9.2$
meV\cite{AgNiO2_mag_order}. Energy scans through the data at
constant wavevectors showing the low energy gap and the full
extent of the scattering are shown in Figs.\
\ref{fig_E1E2E3_Simulation_results}A and E (solid points). The
dashed line in those plots indicates the estimated non-magnetic
background, obtained by a smooth interpolation of the signal
observed at energies below and above the magnetic signal. The
energy gap to magnetic excitations fills up upon heating through
the N\'{e}el temperature [compare Fig.\
\ref{fig_IN6_MARI_mslice}c) with d) and scans shown in Fig.\
\ref{fig_Energy_gap_IN6_with_Temp}]. The structured band of
magnetic scattering is replaced as expected by an overdamped
signal in the paramagnetic phase at high temperatures, compare
Fig.~\ref{fig_IN6_MARI_mslice}a) with b) and various scans in
Fig.~\ref{fig_E1E2E3_Simulation_results}E.

At the lowest temperatures the boundaries of the magnetic
scattering show considerable structure which can be used to impose
constraints on the underlying dispersion relation. In particular
information is contained in the gap, extent of bandwidth, slope of
low-$Q$ dispersion up to the first minimum gap [see Fig.\
\ref{fig_IN6_MARI_mslice}a)] and the low-energy dispersion ($\sim
10\%$ of bandwidth) between subsequent minimum gap wavevectors
[see Fig.~\ref{fig_IN6_MARI_mslice}c)]. The powder data is a
spherical average of the dispersions along all directions in
reciprocal space weighted by the neutron structure factor, so the
low-energy boundary of the powder data corresponds to the
dispersion along some direction in reciprocal space which has the
minimum energy at a given $Q$. In general the spin-wave dispersion
has global minima at magnetic Bragg wavevectors and here strong
low-energy scattering is expected. This is consistent with the
data in Fig.~\ref{fig_IN6_MARI_mslice}c) showing clear lobes of
magnetic scattering intensity coming down in energy near
wavevectors 0.7 and 0.9 \AA$^{-1}$ where the first two magnetic
Bragg peaks (1/2,0,0) and (1/2,0,1) occur; a constant energy scan
near the gap minimum is shown in Fig.\
\ref{fig_E1E2E3_Simulation_results}D. Inelastic intensity occurs
near the same wavevectors as magnetic Bragg peaks in the
diffraction pattern shown in Fig.\
\ref{fig_E1E2E3_Simulation_results}D bottom panel [data from the
GEM diffractometer at ISIS]. The low-energy boundary of the
scattering is clearly modulated as a function of wavevector and
the onset energy varies as a function of wavevector $Q$ as
illustrated by comparing energy scans in Fig.\
\ref{fig_E1E2E3_Simulation_results}B and C.

\section{Analysis}
\label{sec_Analysis}
The modulations observed in the lower boundary of the magnetic
scattering indicate strongly dispersive excitations, however in
principle it is not possible to extract precise dispersion
relations from the powder data, which represents an average along
all directions in reciprocal space. Nevertheless, if the
dispersion is highly asymmetric along different directions and if
in-plane and out-of-plane lattice parameters are sufficiently
different then the lower-boundary of the scattering can be
identified over certain $Q$-ranges with the dispersion along
specific directions in reciprocal space, so one may impose certain
constraints on the dispersion model. Following this approach we
have found two models for the dispersion relation that could both
explain the observed wavevector-dependence of the boundaries of
the powder-averaged spectrum and present them in the following.
\begin{figure}[tbh]
\begin{center}
\includegraphics[width=8cm, bbllx=49,bblly=242,bburx=546,bbury=589,
 angle=0,clip=]{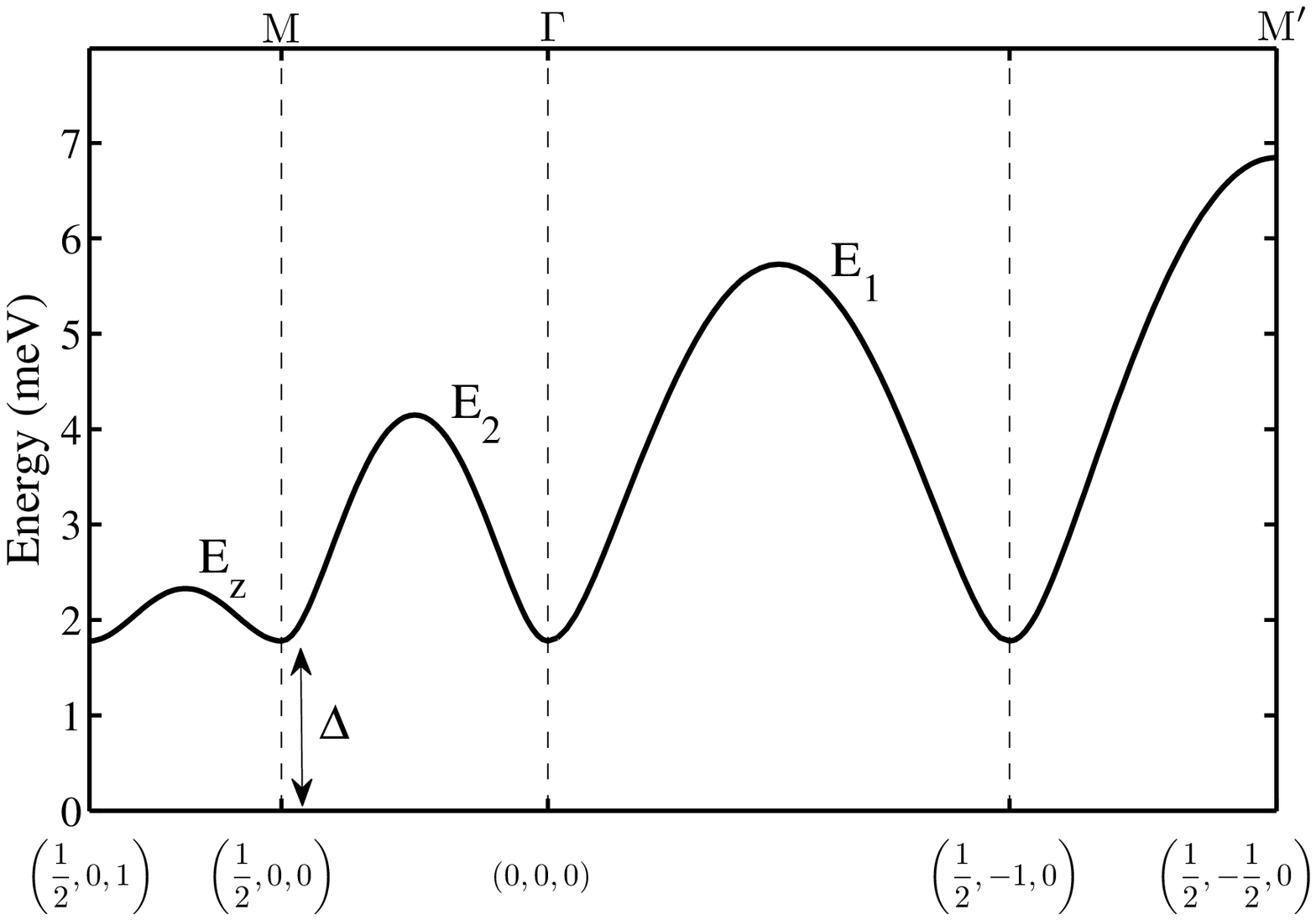}
 \caption{ \label{fig_Panneled_Dispersion}
Sinusoidal dispersion relation in eq.\ (\ref{Eq:dispersion})
plotted along symmetry directions in the Brillouin zone in Fig.\
\ref{fig_BZ}b) (dashed lines). Dispersion parameters are
$\Delta=1.78$ meV, $E_1$=5.73 meV, $E_2=4.15$ meV and $E_z=2.33$
meV.}
\end{center}
\end{figure}
\begin{figure}
\begin{center}
\includegraphics[width=6.11cm,  bbllx=252,bblly=457,bburx=520,  bbury=665]
{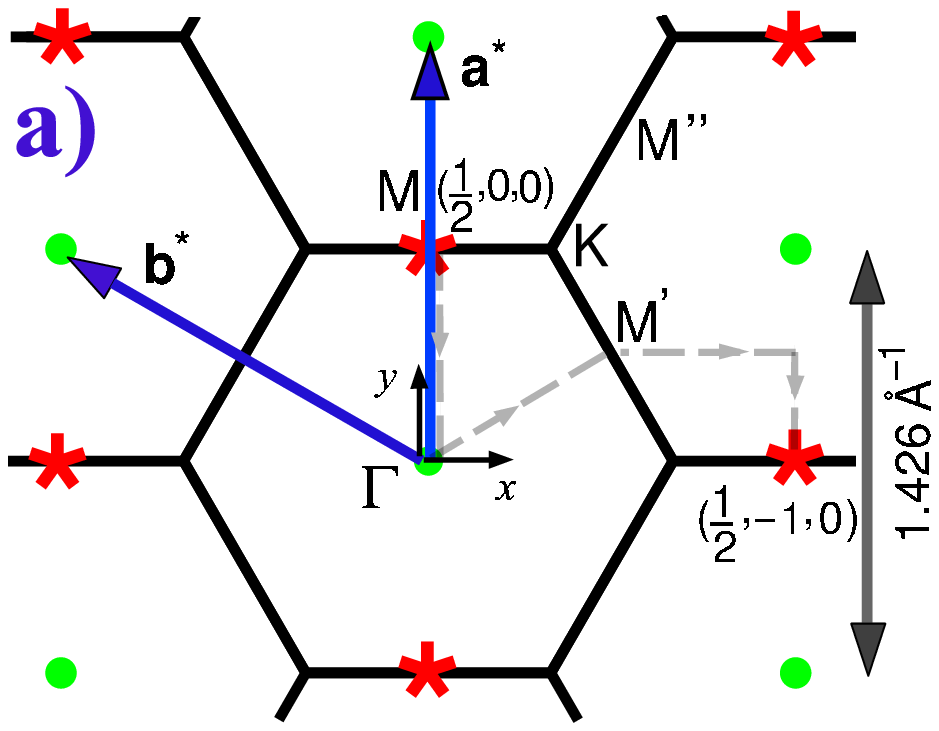} 
\includegraphics[width=6.79cm,  bbllx=0,bblly=14,bburx=435,  bbury=477]
{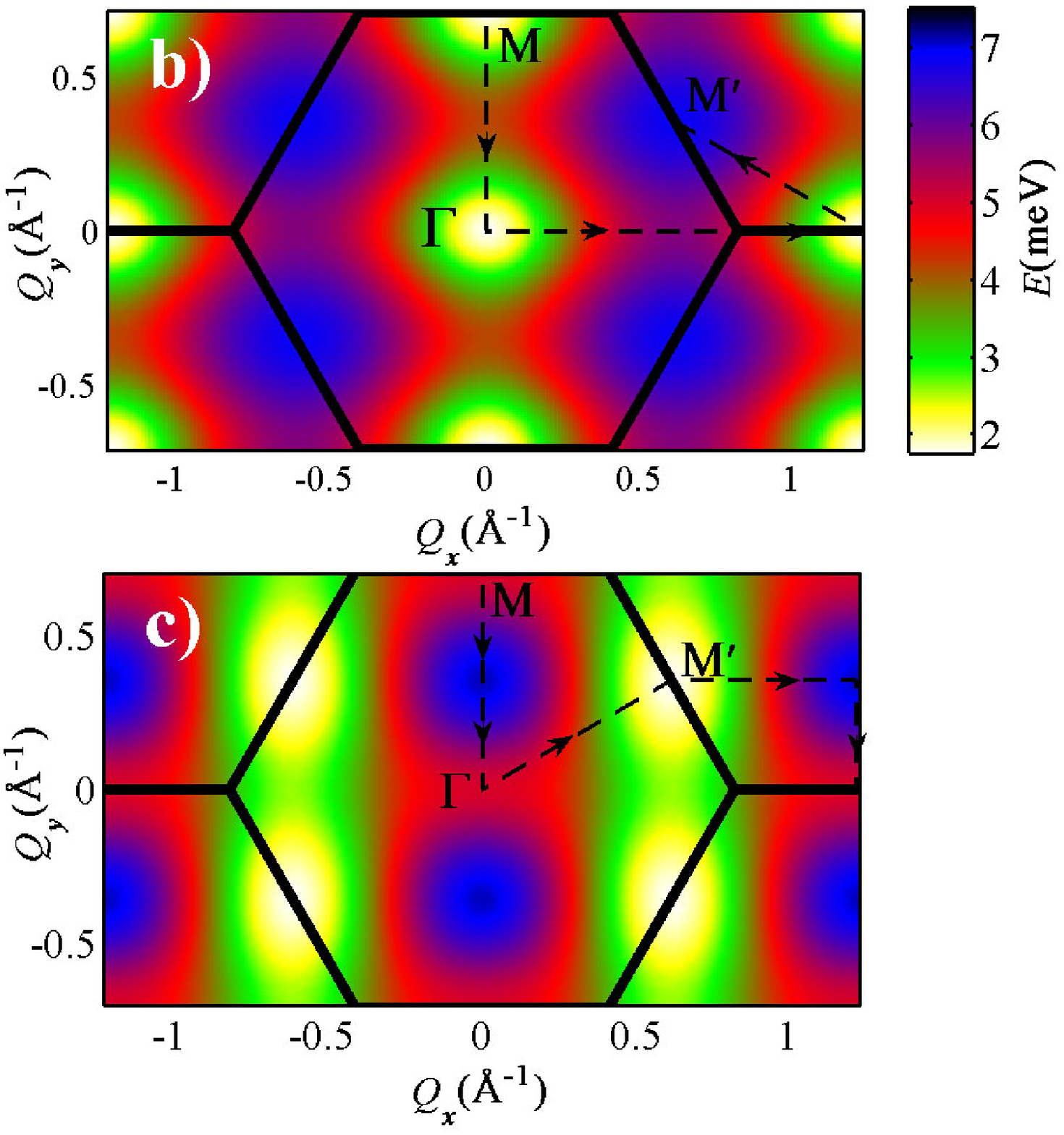}
  \caption{\label{fig_BZ}(Color online) a) Reciprocal basal plane
  of $2H$-AgNiO$_2$ showing the Brillouin zone edges (thick line hexagons),
  zone centers (solid points), and locations (stars) of magnetic
  Bragg peaks from the ordered domain shown in Fig.~\ref{fig_Agnio2_mag_structure}a)
  with stripes along the $\bm{b}$-axis.
  Labels $\Gamma$, M, K indicate special symmetry points in
  the hexagonal Brillouin zone (M$^{\prime}$ and M$^{\prime\prime}$ are
  Bragg peak positions for equivalent ordered domains with stripes
  along $\bm{a}+\bm{b}$ and $\bm{a}$, respectively). Dashed gray lines
  shows directions along which the dispersion
  is plotted in Fig.~\ref{fig_Panneled_Dispersion_J1J2}. Arrows
  labelled $x$, $y$ indicate the natural orthogonal coordinate system
  used for the empirical dispersion model in eq.\ (\ref{Eq:dispersion}).
  b,c) Color contour maps of the dispersion in ($ab$) plane in the two
  models consistent with the data: b) the empirical parameterization
  with a sinusoidal dispersion plotted in Fig.~\ref{fig_Panneled_Dispersion} and
  c) the linear spin wave model for $J_1-J_2$ triangular layers
  with easy axis and inter-layer couplings in Fig.\
  \ref{fig_Panneled_Dispersion_J1J2}d)(dashed line). Note that in model b)
  the minimum gap occurs at the magnetic Bragg
  wavevector M, whereas in model c) the minimum gap occurs at the soft
  point M$'$.}
  \end{center}
\end{figure}

\subsection{Parameterization by an empirical sinusoidal
dispersion model}
\label{sec_sinusoidal}
We first consider a parameterization of the data by an empirical
sinusoidal dispersion model with a finite gap $\Delta$ above the
magnetic Bragg peaks and different zone boundary energies along
the three orthogonal directions in the Brillouin zone: $E_1$ along
the magnetic stripe direction, $E_2$ transverse to stripes in the
plane, and $E_z$ in the inter-layer direction. In detail we
consider the following {\em empirical form} for the dispersion
relation $\omega_{\bm Q}$
\begin{eqnarray}
\label{Eq:dispersion} (\omega_{\bm
Q})^{2}=&\Delta^2+(E_1^2-\Delta^2)~\sin^2(Q_x a/2)+\nonumber \\
&+(E_2^2-\Delta^2) ~\sin^2(Q_y a \sqrt{3}/2)+\nonumber \\
&+(E_z^2-\Delta^2) ~\sin^2(Q_z c/2)
\end{eqnarray}
where $Q_x$, $Q_y$, $Q_z$ are components (in \AA$^{-1}$) of the
wavevector transfer ${\bm Q}$ in an orthogonal coordinate system
with $x$ along the magnetic stripes, $y$ transverse to stripes in
plane, $z$ normal to the planes. The transformation to wavevector
components ($h,k,l$) in units of the reciprocal lattice of the
hexagonal unit cell is given by
\begin{eqnarray}
\label{Eq:rlu2xyz} Q_x&=&-\frac{2\pi}{a}k, \nonumber \\
Q_y & =& \frac{2\pi}{a\sqrt{3}}\left(2 h+k \right), \nonumber\\
Q_z &= &\frac{2\pi}{c}l.
\end{eqnarray} Fig.~\ref{fig_Panneled_Dispersion} shows
a plot of this dispersion relation along various symmetry
directions in the Brillouin zone and Fig.~\ref{fig_BZ}b) shows a
2D contour map of the dispersion. The minimum gap $\Delta$ is
reached at the origin and at all magnetic Bragg peak positions
$(H,K,L)\pm \bm{k}$ with $\bm{k}=(1/2,0,0)$ the ordering
wavevector and $H$, $K$ and $L$ integers. In order to compare this
dispersion relation with the data one needs also a model for the
neutron scattering cross-section. We assume the following simple
form for the one-magnon cross-section
\begin{eqnarray}
S^{xx}({\bm Q},\omega)& =&  S^{yy}({\bm Q},\omega) \nonumber \\
 & =& \mathcal{C}\frac{S}{2} \left(\omega_{\bm{k}}\right)_{\rm max}
\frac{1-\gamma_{\bm Q}}{\omega} ~ G(\omega-\omega_{\bm Q})
\label{Eq:E1E2E2_Sqw}
\end{eqnarray}
Here $\mathcal{C}$ is an overall scale factor and $(1-\gamma_{\bm
Q})$ is a geometric factor to concentrate the intensity near the
antiferromagnetic Bragg peak wavevectors ($H\pm1/2,K,L$) with $H$,
$K$, $L$ integers, 
$\gamma_{\bm Q}=\cos(Q_x a/2)\cos(Q_y a \sqrt{3}/2)$. The
$1/\omega$ factor is a characteristic energy-dependence of the
spin-wave structure factor for antiferromagnets and
$\left(\omega_{\bm{k}}\right)_{\rm max}$ is the maximum magnon
energy for all wavevectors $\bm{k}$, and is introduced here for
dimensionality purposes. $S=1$ is the spin quantum number. Eq.\
(\ref{Eq:E1E2E2_Sqw}) includes the fact that spin fluctuations
occur in the plane transverse to the ordered spin direction in a
collinear antiferromagnet, so $x$ and $y$ refer to two orthogonal
directions in the hexagonal $ab$ plane and $z$ is along $c$. This
form is a generic structure factor for many simple spin-wave
models of collinear antiferromagnets. Here $G(\omega-\omega_{\bm
Q})$ is a gaussian function which models the instrumental
resolution.

A spherical average of the intensity distribution of eq.\
(\ref{Eq:E1E2E2_Sqw}) including the neutron polarization factor,
magnetic form factor for Ni$^{2+}$ ions [as in eq.\
(\ref{eq_App_cross_section})] and an estimated additive
non-magnetic background (including the incoherent elastic line)
was compared to the data and the overall comparison is shown in
the middle panels of Fig.~\ref{fig_IN6_MARI_mslice}. The overall
bandwidth and the observed structure in the data at low energies
is well reproduced by this model for gap $\Delta=1.78(5)$ meV,
$E_1$=5.73(4) meV, $E_2=4.15(6)$ meV and $E_z=2.33(5)$ meV in eq.\
(\ref{Eq:dispersion}). The maximum magnon energy is
$\left(\omega_{\bm{k}}\right)_{\rm max}=7.0$ meV and an overall
scale factor $\mathcal{C} \simeq 0.35$ in eq.\
(\ref{Eq:E1E2E2_Sqw}) gives intensities comparable to data.

\begin{figure}[tbh]
\begin{center}
\includegraphics[width=7.5cm,  bbllx=0,bblly=0,bburx=245,  bbury=310]
 {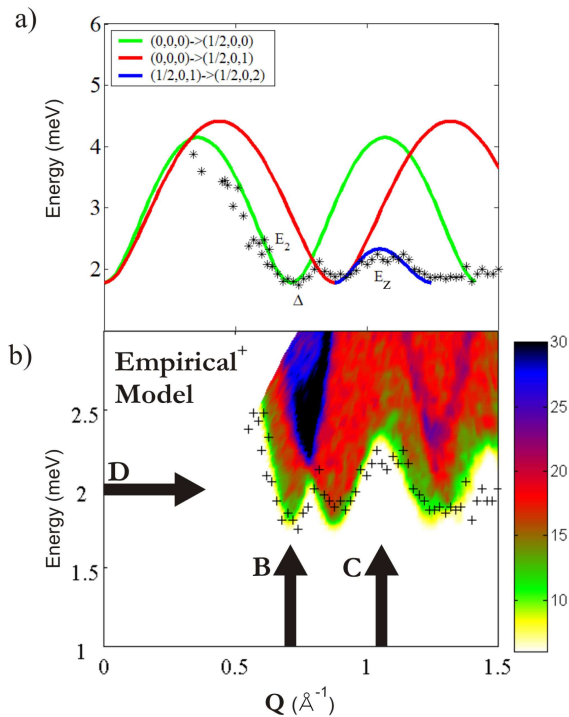}
\caption{\label{fig_E1E2E3_dispersion_and data} (Color online)
  a) Dispersion relations for the empirical sinusoidal model in eq.\
  (\ref{Eq:dispersion}) along various directions involved in defining
  the low-energy edge of the powder-averaged spectrum.
  b) Corresponding powder-averaged spectrum.
  Stars in both panels are the experimentally-determined low-energy edge of
  the magnetic scattering. Thick arrows indicate location of
  scans B,C, D in Fig. \ref{fig_E1E2E3_Simulation_results}.}
 \end{center}
\end{figure}

To analyze the agreement quantitatively we first show in Fig.\
\ref{fig_E1E2E3_dispersion_and data}a) how the lower boundary of
the powder-averaged magnetic signal can be identified over certain
wavevector ranges with the dispersion along particular directions
in reciprocal space. In particular the observed sharp low-$Q$
dispersion up to the first gap minimum near $Q_{\rm M}=0.7$
\AA$^{-1}$ in Fig.~\ref{fig_IN6_MARI_mslice}a) is attributed to
the dispersion between the origin $\Gamma$(000) and the closest
magnetic Bragg peak at M(1/2,0,0); scan A along the direction
indicated by the bold arrow in Fig.~\ref{fig_IN6_MARI_mslice}a) is
therefore very sensitive (at low energies) to the $\Gamma$M
dispersion and a fit to this scan shown in Fig.\
\ref{fig_E1E2E3_Simulation_results}A gives $E_2=4.15 \pm 0.06$
meV, the zone boundary along the direction transverse to stripes
in plane. To explain the observed full extent of the bandwidth up
to 7.5 meV another dispersion, larger in magnitude is required and
this is assigned to the direction along the stripes, $E_1$, since
the inter-layer dispersion bandwidth $E_z-\Delta$ is assumed to be
small. Quantitative fits give $E_1=5.73 \pm 0.04$ meV. The energy
scan in Fig.~\ref{fig_E1E2E3_Simulation_results}B is sensitive to
the gap above the magnetic Bragg wavevector, obtained as
$\Delta=1.78(5)$ meV. The energy scan in Fig.\
\ref{fig_E1E2E3_Simulation_results}C is at a wavevector
corresponding to the midpoint between two subsequent Bragg
wavevectors along the $c$-axis, (1/2,0,1) and (1/2,0,2) [see Fig.\
\ref{fig_E1E2E3_dispersion_and data}a)], so this scan is sensitive
to the inter-layer zone boundary and fits give $E_z=2.33 \pm 0.05$
meV, thus obtaining the last parameter of the empirical dispersion
relation.

Fig.~\ref{fig_E1E2E3_dispersion_and data} shows in more detail
the agreement between the experimentally-determined onset of the
magnetic scattering (stars) and the empirical model (various solid
lines) in panel a) and the calculated powder-averaged spectrum in
panel b). The onset points were determined from scans through the
measured powder data at fixed wavevector $Q$ and locating the
energy where the magnetic intensity was above a minimum threshold
value. The model reproduces well the shape of the lower edge of
the scattering. Fits to specific scans shown in Fig.\
\ref{fig_E1E2E3_Simulation_results}A-D give good account of the
onset and upper boundary of the scattering. The form of the
intensity distribution within the excitation band is not
replicated in detail, as seen for example in Fig.\
\ref{fig_E1E2E3_Simulation_results}A where the simulation gives a
more pronounced trough in the intensity in the middle of the band.
This is almost certainly due to the over simplification of the
model.

\begin{figure*}[tbh]
\begin{center}
\includegraphics[width=6.56cm, bbllx=69,bblly=308,bburx=468,bbury=566,
 angle=0,clip=]{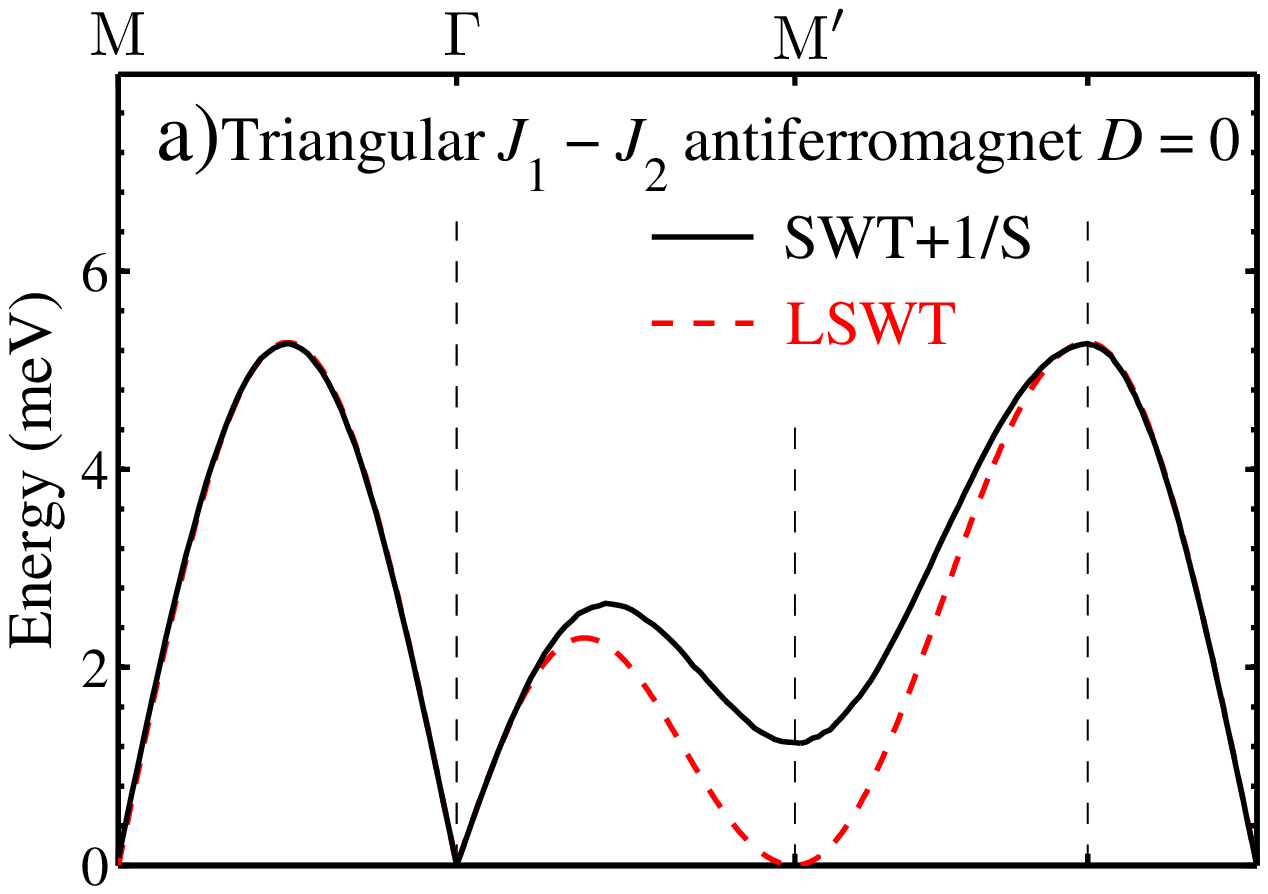}\hspace{0.25cm}
\includegraphics[width=7.7cm, bbllx=69,bblly=308,bburx=540,bbury=566,
 angle=0,clip=]{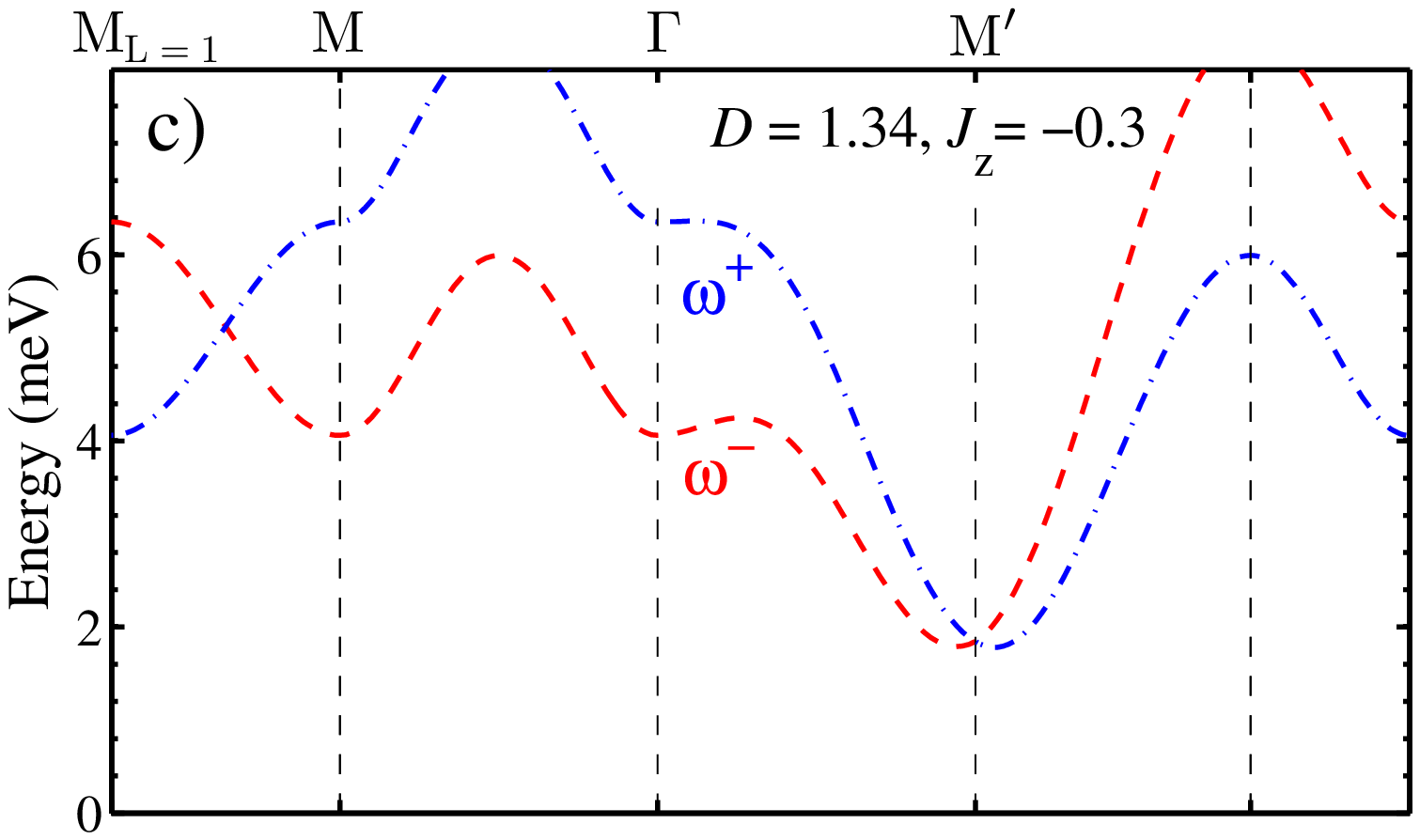}\hspace{-2mm}
 \includegraphics[width=2.13cm, bbllx=81,bblly=307,bburx=212,bbury=573,
 angle=0,clip=]{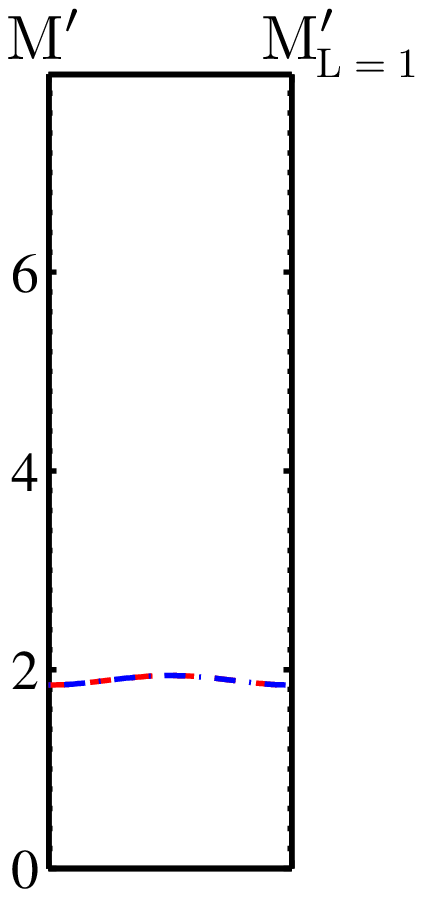}
 \includegraphics[width=6.56cm, bbllx=69,bblly=275,bburx=468,bbury=542,
 angle=0,clip=]{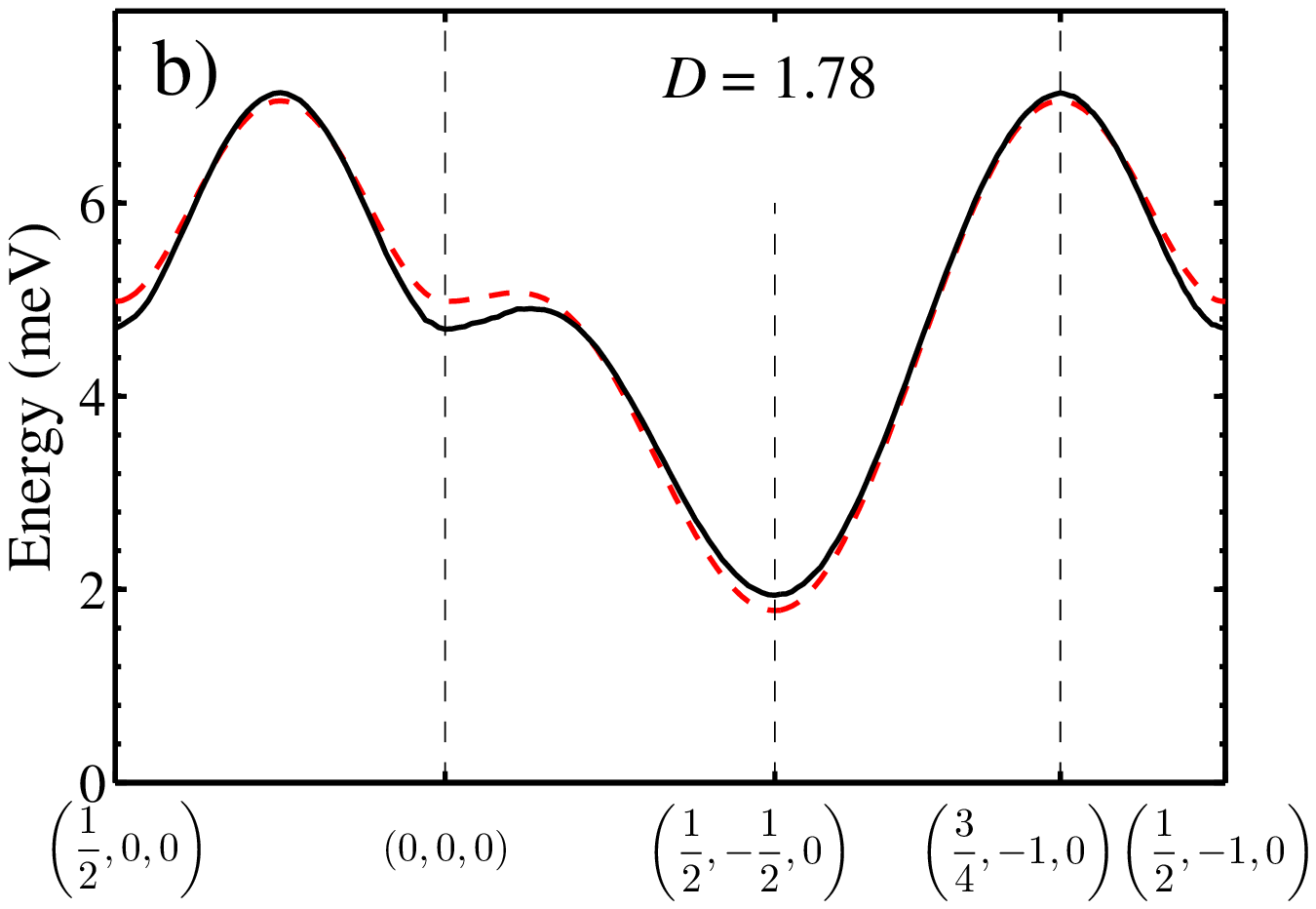}\hspace{0.25cm}
\includegraphics[width=7.7cm, bbllx=69,bblly=276,bburx=540,bbury=542,
 angle=0,clip=]{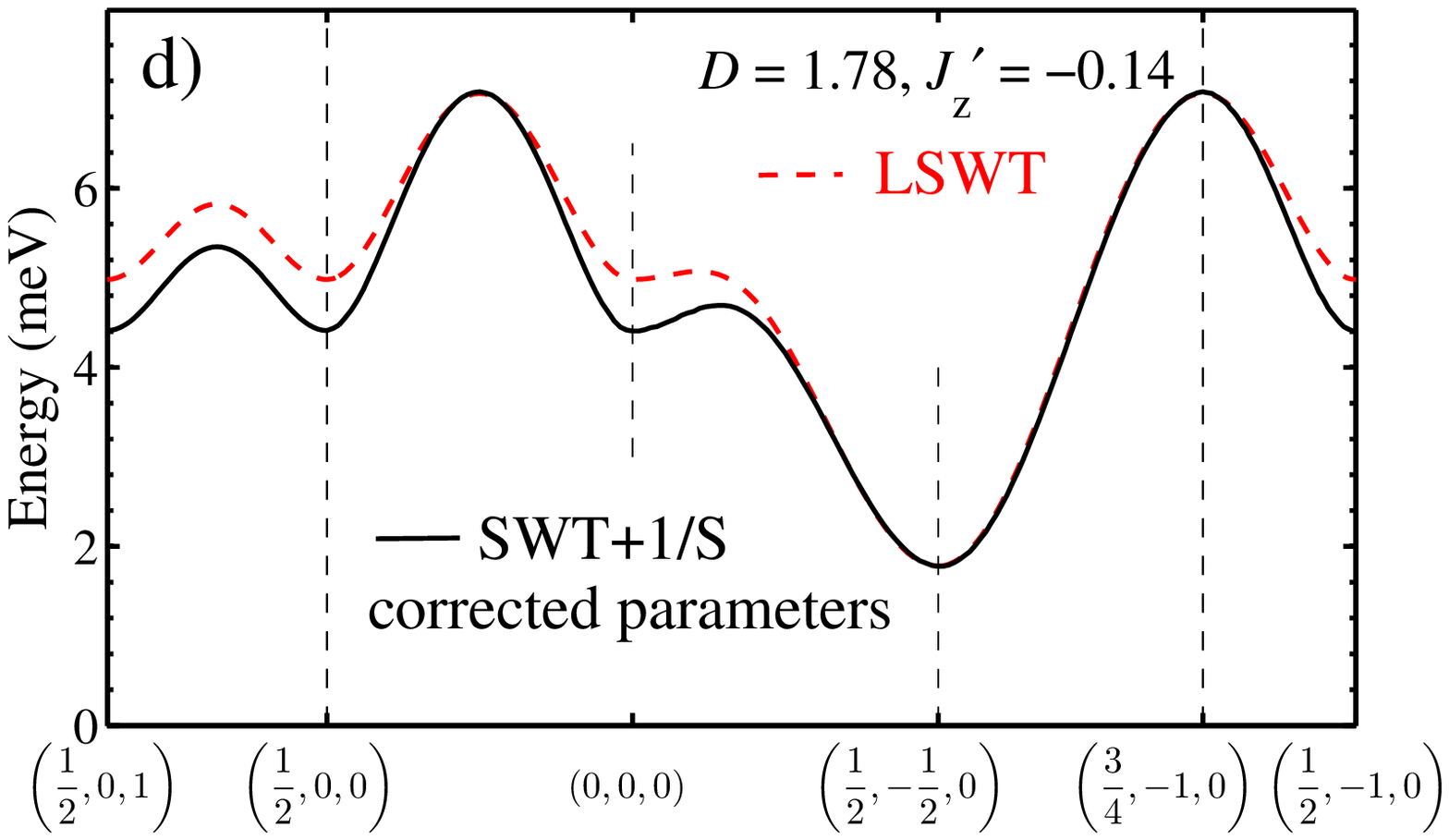}\hspace{-2mm}
\includegraphics[width=2.13cm, bbllx=81,bblly=276,bburx=212,bbury=542,
 angle=0,clip=]{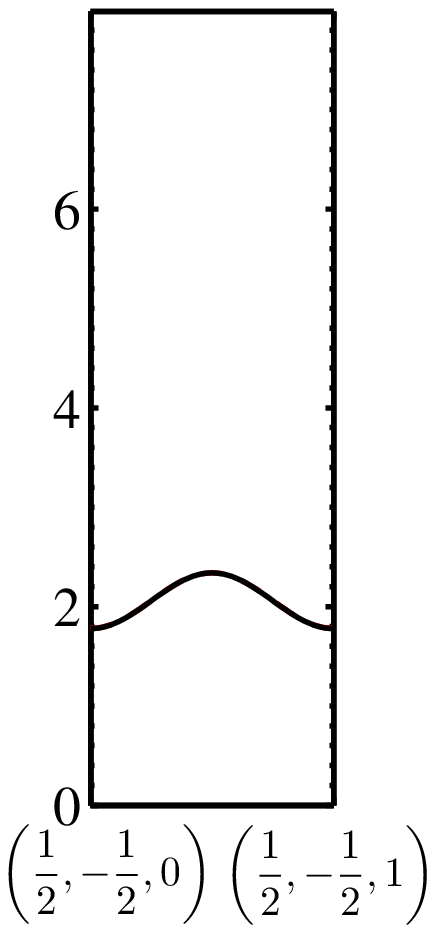}
\caption{ \label{fig_Panneled_Dispersion_J1J2} (Color online)
Dispersion relation for the stacked triangular antiferromagnetic
layers within LSWT (dashed lines) and SWT+$1/S$ (solid lines) for
various values of the parameters. The dispersion is plotted along
high-symmetry directions in reciprocal space (dashed line path in
Fig.~\ref{fig_BZ}a), top labels are high symmetry points in the
Brillouin zone. a) Triangular lattice with antiferromagnetic
couplings $J_1=1.32$ meV, $J_2=0.15J_1$ and no anisotropy. $1/S$
quantum corrections have a large effect at the soft point M$'$. b)
Same as a) but with easy-axis anisotropy ${\cal D}=1.78$ meV,
$1/S$ corrections are much smaller. c) Triangular layers with
easy-axis anisotropy and coupled by NN interlayer exchange
$J_z=-0.3$ meV. This leads to two incommensurate spin-wave modes
with very little inter-layer dispersion near the gap minimum (last
panel in the row), not consistent with data. d) Triangular layers
with anisotropy coupled by second-layer exchange $J_z'=-0.14$ meV,
this model gives the best fit to the data. Dashed (solid) line
shows best fit using LSWT (${\rm SWT}+1/S$) with parameter values
given in Table~\ref{fit_table}.}
\end{center}
\end{figure*}

\begin{figure*}
\begin{center}
\includegraphics[width=7.5cm, bbllx=0,bblly=0,bburx=333,
bbury=250] {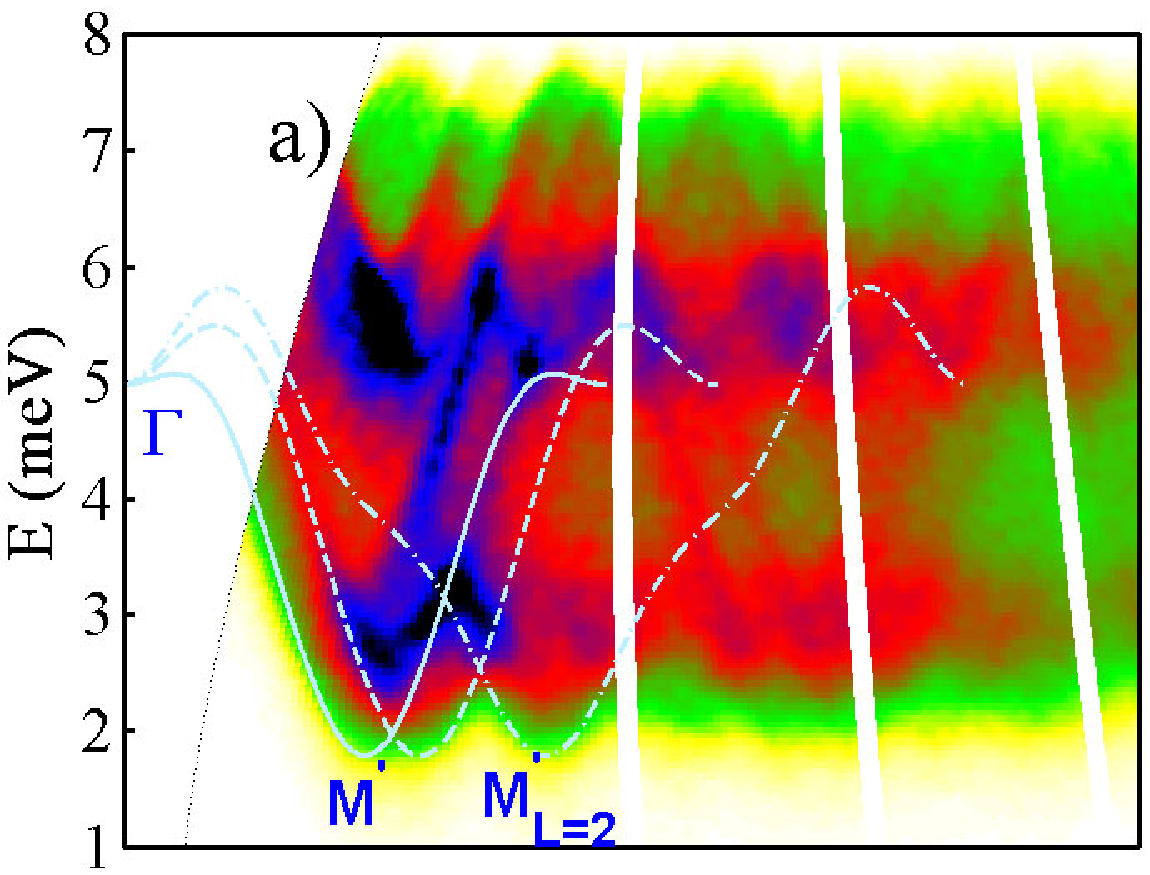} \hspace{0.3cm}
\includegraphics[width=8.0cm, bbllx=0,bblly=0,bburx=354,
bbury=249] {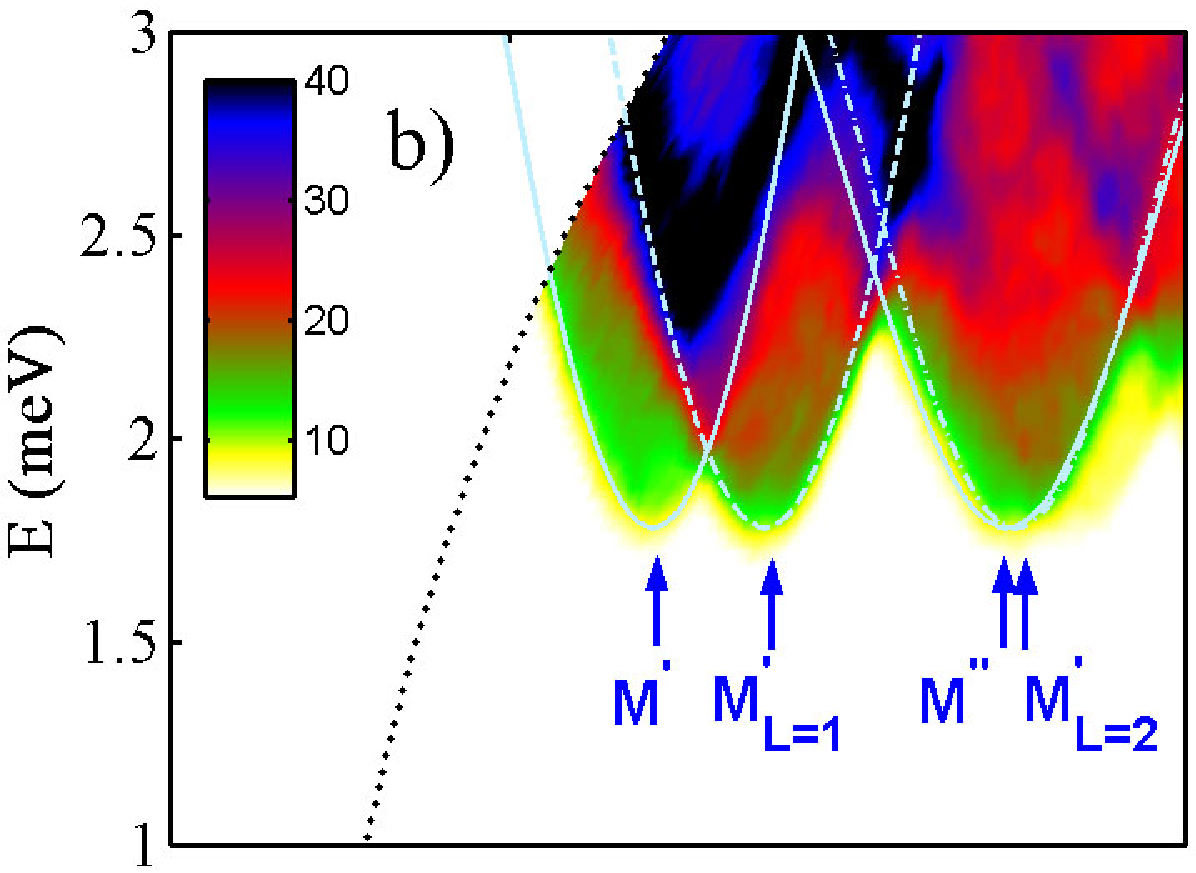}
\includegraphics[width=7.5cm, bbllx=0,bblly=0,bburx=331,
bbury=275] {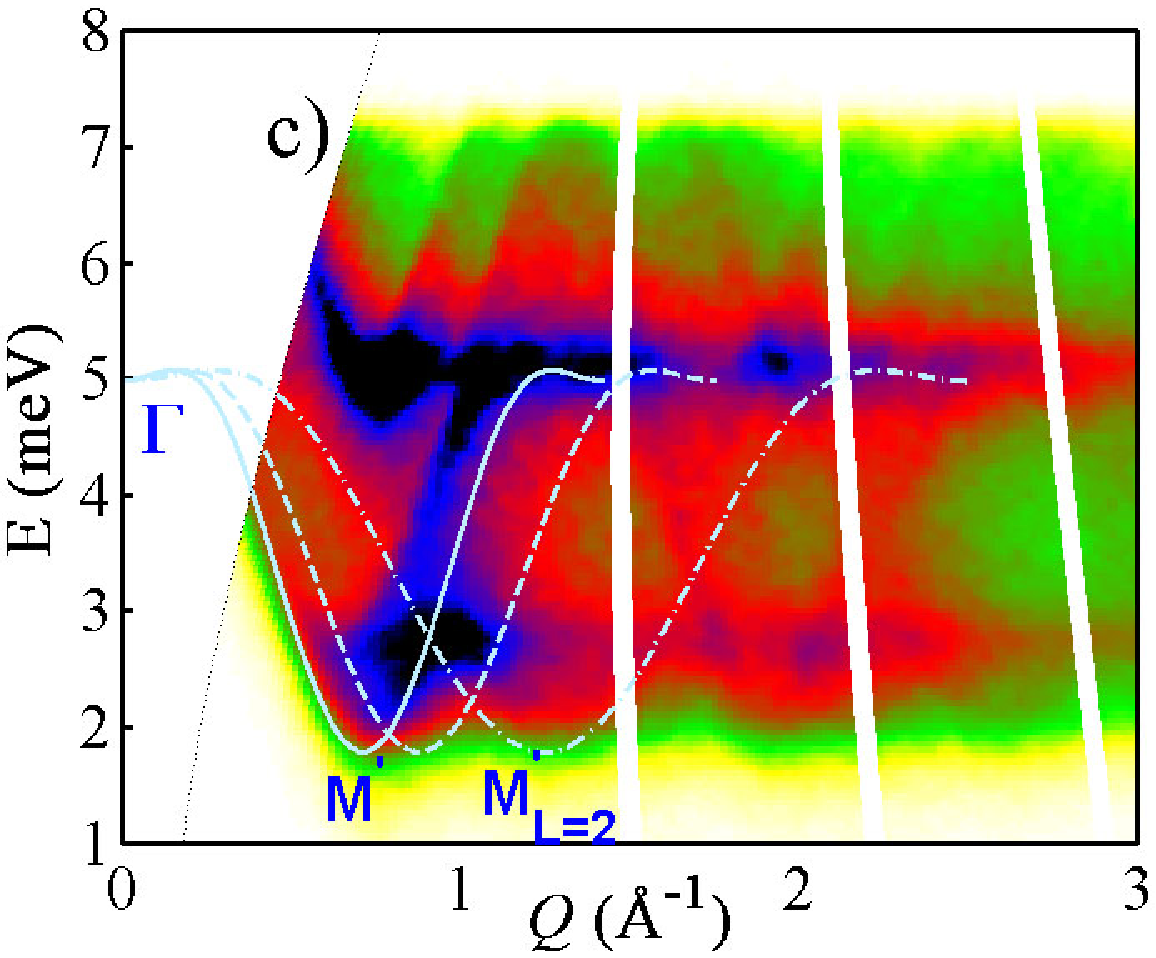} \hspace{0.3cm}
\includegraphics[width=8.0cm, bbllx=0,bblly=0,bburx=354,
bbury=275] {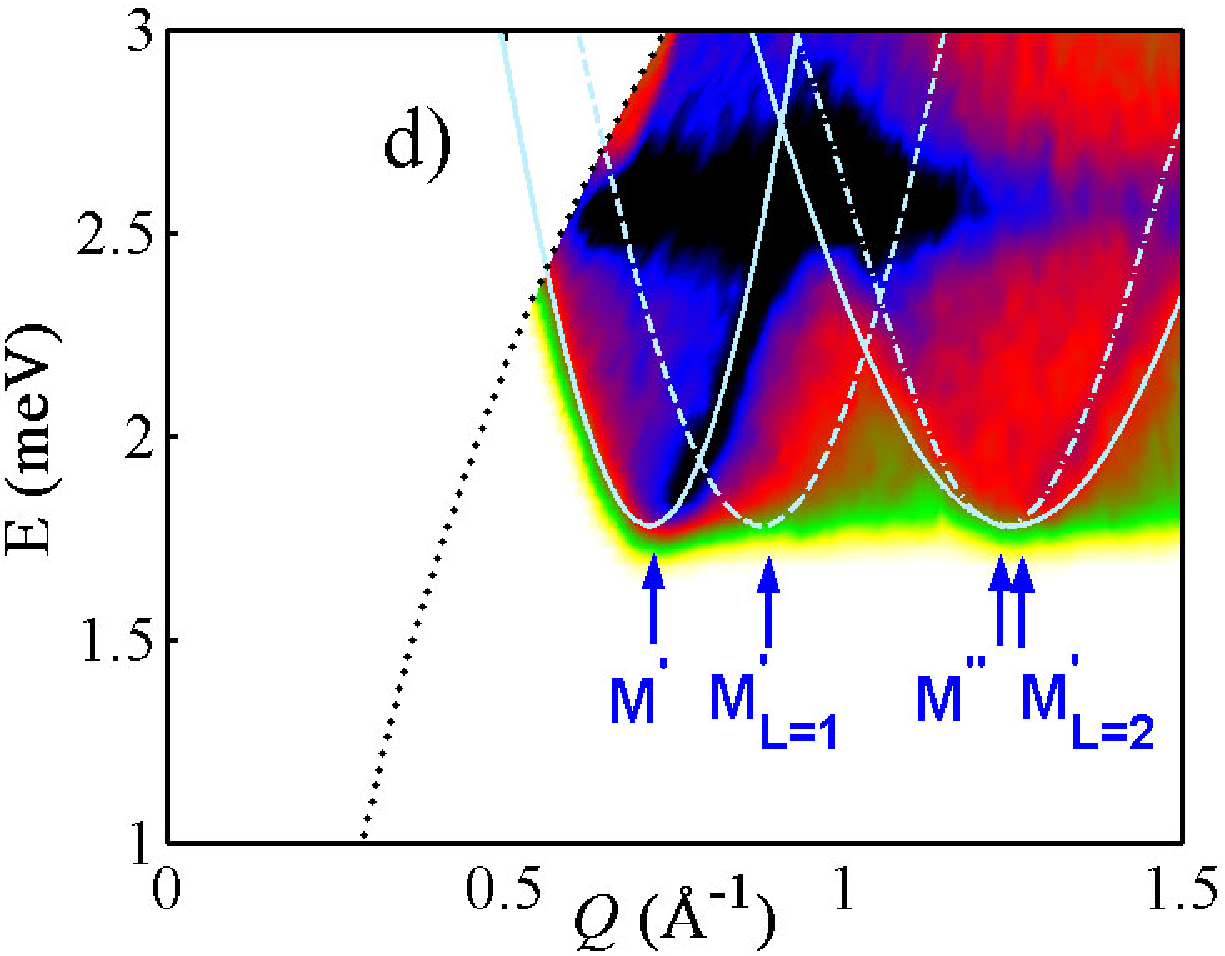}
  \caption{\label{fig_lswt}(Color online) Powder averaged
  spin-wave spectrum for easy-axis triangular layers with/without
  (top/bottom) interlayer couplings. Left panels
  to be compared with Fig.~\ref{fig_IN6_MARI_mslice}a)
  and right panels to Fig.~\ref{fig_IN6_MARI_mslice}c). The
  parameters are $J_1=1.32$\ meV, $J_2=0.15 J_1$, ${\cal D}=1.78$\ meV,
  $J_z=0$, and $J_z'=-0.14$\ meV (top panels) or $J_z'=0$\ meV (bottom
  panels). The calculations contain an additive flat background. The
  various lines overplotted onto the intensity color maps show the dispersion
  along the directions from the origin $\Gamma$ point to the
  closest soft points at M$^{\prime}$(1/2,-1/2,0)
  (and the $L=1$ and $L=2$ versions of this) and
  M$^{\prime\prime}$(1,-1/2,0), with minima at wavevectors indicated
  by the vertical arrows in the right panels. The
  dispersion at the lower boundary of the scattering in b) is very
  similar to the data in Fig.~\ref{fig_IN6_MARI_mslice}c) and are
  attributed to the weak interlayer couplings (no interlayer dispersion is
  present in d) calculated for decoupled layers). The dotted lines
  at low-$Q$ mark the left edge of the region probed by the experiments.}
  \end{center}
\end{figure*}
\begin{figure*}
\begin{center}
\includegraphics[width=16cm, bbllx=20,bblly=334,bburx=539,
bbury=526] {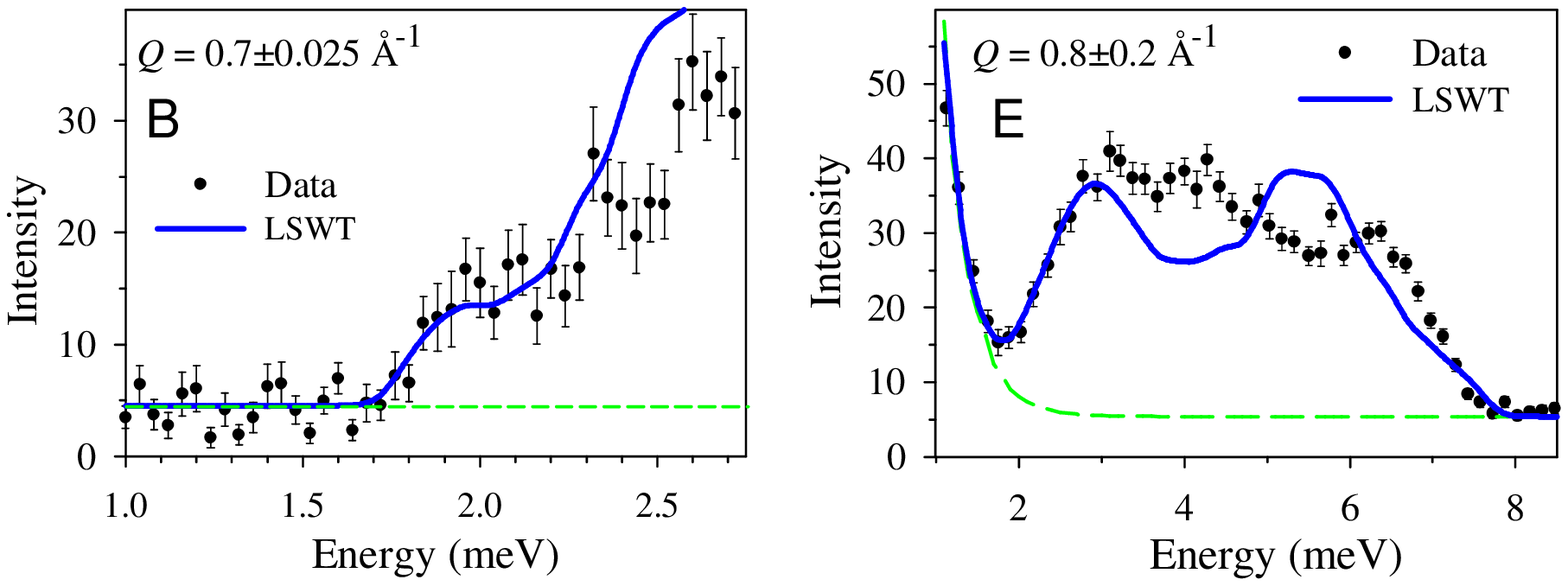}
  \caption{\label{fig_lswt_fits}(Color online) Comparison
  between data (filled points) and the linear spin-wave model
  for easy-axis triangular layers (solid lines). Dashed
  lines show the estimated non-magnetic background and parameters
  are as in Fig.\ \ref{fig_lswt}a).}
  \end{center}
\end{figure*}
\subsection{Parameterization in terms of an easy-axis Heisenberg model on a  triangular lattice
} \label{sec_coupled_layers}
The empirical sinusoidal dispersion model considered in the
previous section is the simplest dispersion model that can be used
to parameterize and explain quantitatively the observed dispersion
of the boundaries of the powder-averaged magnetic inelastic
scattering. This analysis suggested that the dispersion relation
is gapped and predominantly two-dimensional. Here we compare the
data quantitatively with predictions of linear spin wave theory
for a microscopic model of coupled triangular layers where we
include antiferromagnetic in-plane couplings for the first and
second-neighbour exchanges $J_1$ and $J_2$, an easy-axis
anisotropy and allow for both first- and second-layer couplings
$J_z$ and $J_z'$, as shown in Fig.~\ref{fig_Structure}a,b).
Specifically, we consider the spin Hamiltonian
\begin{eqnarray}
{\cal H}_0 &= \sum_{{\rm NN}} J_1 \bm{S}_{i}\cdot \bm{S}_{j}+
\sum_{{\rm{NNN}}} J_2 \bm{S}_{i} \cdot \bm{S}_{l} \nonumber \\
& -{\cal{D}}\sum_{i}\left(S^{z}_{i}\right)^2 +\sum_{{\rm
InterlayerNN}} J_{z} \bm{S}_{i} \cdot \bm{S}_{k} \nonumber \\
& +\sum_{{\rm Interlayer NNN}} J_{z}' \bm{S}_{i} \cdot \bm{S}_{n}
, \label{eq:Hamiltonian}
\end{eqnarray}
where the spin is $S=1$, NN and NNN stand for nearest-neighbour
and next-nearest-neighbour and each spin pair is counted once in
the summation. The third term is an easy axis anisotropy, needed
to select the ordering spin direction along $c$, and to generate a
spin gap in the spectrum. In the absence of a clear physical
picture of the origin of the anisotropy, we have chosen a
single-ion form for this term. We anticipate that an exchange
anisotropy $(\delta J)^z S_i^z S_j^z$ would have qualitatively
similar effects. Both in-plane exchanges are antiferromagnetic
$J_1,J_2>0$ whereas the inter-layer couplings are taken to be
ferromagnetic $J_{z},J_z'<0$, to stabilize the observed magnetic
structure shown in Fig.~\ref{fig_Spin_wave_sketch}.

We have calculated the spin-wave dispersion and dynamical
correlations for the spin Hamiltonian (\ref{eq:Hamiltonian}) in
the large $S$-limit using the standard Holstein-Primakoff
formalism and details are given in Appendix~\ref{sec_appendix}
with the dispersion relations in eq.~(\ref{eq_App_dispersion}). To
illustrate the effect of the various terms in the Hamiltonian we
plot the dispersion for a triangular lattice with no anisotropy in
Fig.~\ref{fig_Panneled_Dispersion_J1J2}a) (dashed line). The
linear spin-wave dispersion has zero modes not only at the
magnetic Bragg peak position M for the specific ordered magnetic
domain, but also at soft points such as M$^{\prime}$ in
Fig.~\ref{fig_BZ} which are Bragg peak positions for the
symmetry-equivalent domains (there are three possible magnetic
domains related by a three-fold rotation around the $c$-axis).

The presence of such unphysical gapless modes is a consequence of the
macroscopic ground state degeneracy at the classical level and
including $1/S$ corrections
[Fig.~\ref{fig_Panneled_Dispersion_J1J2}a) solid line] generates a
gap at the soft points but preserves a gapless Goldstone mode at
the Bragg wavevectors.\cite{Chubukov_1992}  The situation is
completely different in the presence of an easy-axis anisotropy
which generates a gap everywhere [see Fig.~\ref{fig_Panneled_Dispersion_J1J2}b)],
the gap at the Bragg
wavevector increases very rapidly as a power law, whereas at the
soft point increases only linearly with anisotropy, so above some
threshold anisotropy value the minimum gap is no longer at the
magnetic Bragg wavevector M, but moves to the soft point M$'$.

Coupling the layers by a NN exchange $J_z<0$ (ferromagnetic) has
the effect of splitting the dispersion into two modes [see
Fig.~\ref{fig_Panneled_Dispersion_J1J2}c)], corresponding to acoustic
and optic magnon modes between the two layers in the unit cell.
The gap minimum is no longer at the commensurate soft point M$'$
but at incommensurate positions symmetrically displaced from M$'$
along the magnetic stripe direction, see
Fig.~\ref{fig_Panneled_Dispersion_J1J2}c).
The incommensurate position
of the gap minimum varies linearly with $|J_z|$ for small $J_z$
and it originates physically from the fact that hopping of magnons
is somewhat frustrated as an up spin in the bottom layer interacts
through $J_z$ with two up and one down spin in the layers above
and below, see Fig.~\ref{fig_Spin_wave_sketch}b). Moreover, there
is very little interlayer dispersion of the gap minimum, i.e.
between soft points M$'$ with different $L$-values [see
Fig.~\ref{fig_Panneled_Dispersion_J1J2}c) last panel], also due to
frustration in the magnon interlayer hopping at those wavevectors.

Both of these features, incommensurate minima in the dispersion
displaced (in-plane) away from M$'$ and very weak inter-layer
dispersion of the gap minimum are inconsistent with the data,
which shows minima in the scattering at wavevectors very close to
those corresponding to the commensurate M$'$ point and a
significant ($\sim10$\% of the total bandwidth) interlayer
dispersion of the gap minimum. This suggests that the observed
interlayer dispersion is not due to the nearest-neighbour
interlayer exchange $J_z$ but is due to another inter-layer
coupling.

To find a possible explanation for the observed inter-layer
dispersion we now assume $J_z=0$ and consider the effects of a
ferromagnetic second-layer coupling $J_z'$ straight up along the
$c$-axis, see Fig.~\ref{fig_Structure}b). The resulting dispersion
is plotted in Fig.\ \ref{fig_Panneled_Dispersion_J1J2}d) dashed
line, there is a single mode with essentially the same features as
for decoupled triangular layers [see
Fig.~\ref{fig_Panneled_Dispersion_J1J2}b)] with global minima at
the commensurate soft point M$'$ but now there is a modulation in
the energy as a function of $L$-value to first order in $J_z'$.
This model can account for all observed dispersions in the
boundaries of the powder-averaged spectrum and the calculation for
the best fit parameter set is plotted in Fig.~\ref{fig_lswt}a-b)
to be compared with data in
Fig.~\ref{fig_IN6_MARI_mslice}$\rm{a-c)}$.

In the calculation we used the dynamical correlations in eq.~(\ref{eq_App_Sqw}),
included polarization factor and magnetic form factor as in eq.\
(\ref{eq_App_cross_section}), and performed a spherical average.
An overall intensity scale factor ${\cal C} \sim 0.57$ gives
intensities comparable to data, we do not attach special
significance to this value except that it is comparable to the
LSWT prediction of ${\cal C}=1$. The intensity distribution is
also reasonably well captured by the model as shown by the typical
scans in Fig.\ \ref{fig_lswt_fits}, any differences compared to
the data may be due to interactions beyond the minimal spin
Hamiltonian considered.

To obtain quantitative values for the model parameters we have
used the fact that various parts of the spectrum are mostly
sensitive to only one parameter, in particular the gap depends
only on anisotropy via $\Delta_{{\rm M}'} =
2S{\cal{D}}\left(1-\frac{1}{2S}\right)$,
 the upper boundary of the spectrum
is mostly sensitive to the main coupling $J_1$, the magnitude of
the low-energy dispersion between the gap minima with $L=1$ and 2
is given by $4S|J_z'|$. The remaining parameter $J_2$ is extracted
from the slope of the low-$Q$ dispersion up to the first gap
minimum. Specifically, a fit to scan B at the minimum gap [Fig.\
\ref{fig_lswt_fits}B] gives the anisotropy ${\cal{D}}$, scan C
probes the inter-layer dispersion so gives $J_z'$, fits to scans A
and E upper boundary [Fig.\ \ref{fig_lswt_fits}E] give the main
coupling $J_1$ and fits to scan A lower boundary give the
secondary coupling $J_2$.

The best fit values for the parameters obtained this way are shown in Table~\ref{fit_table}.
The calculated powder-averaged spectrum is shown in
Fig.~\ref{fig_lswt}a-b) and compares well with the data in
Fig.~\ref{fig_IN6_MARI_mslice}a-c). To emphasize the sensitivity
of the data to the 3D couplings we also plot in Fig.\
\ref{fig_lswt}c-d) the powder-averaged spectrum assuming decoupled
layers, which reproduces the bandwidth, gap and slope of the
low-$Q$ dispersion up to the first gap minimum but not the
low-energy dispersion ($\sim10\%$ of bandwidth) between subsequent
minimum gap wavevectors.

In the above analysis we neglected the NN interlayer coupling
$J_z$ since the powder data is not very sensitive to its presence,
in particular a finite $J_z$ produces only a negligible
inter-layer dispersion of the gap minimum (dispersion appears only
at order $J^2_z/{\cal D}$ for small $J_z$). Using the fact that a
finite $J_z$ produces a small incommensurate shift of the minimum
gap wavevector an upper bound for the maximum $J_z$ that would
still be consistent with the data can be estimated as
$|J_z|\rlap{\lower4pt\hbox{\hskip1 pt$\sim$}}
\raise1pt\hbox{$<$}~0.25$\ meV.

Within the minimal model we considered in eq.~(\ref{eq:Hamiltonian}) the
estimated values of the interaction parameters would give a Curie-Weiss constant
expected in a high-temperature susceptibility measurement on a powder sample
as\cite{Yoshida1951}
$$
k_{\rm B} \theta = -\frac{1}{3}S(S+1)(6J_1+6J_2+2J_z')= -5.9 {\rm meV}
$$
This is smaller than, but comparable to, the experimental value of $k_{\rm B}\theta=-9.2$ meV
extracted from high-temperature susceptibility
data\cite{AgNiO2_mag_order} and the difference may be due to other
interactions beyond the minimal model we have considered to
explain the dispersive boundaries of the powder data.

\begin{table}
\caption{
Table of best fit values for easy-axis Heisenberg model applied to AgNiO$_2$.
In both cases fits are obtained by comparing the spin wave dispersion calculated
within linear spin wave theory, and including leading interaction corrections,
with the envelope of the powder-averaged spin structure factor $S(Q,E)$
measured by inelastic neutron scattering.    We set $S=1$ and $J_z \equiv 0$ in all fits.
All exchange parameters are given in meV.}
\begin{ruledtabular}
\begin{tabular}{c|c|c|c|c|c|c|c|c|c|c|c|c|c|c|c|c}
\label{fit_table}
                      &   $S$  & $J_1$ & $J_2$ & ${\cal D}$ & $J_z$ & $J_z'$\\
\hline
LSWT              & 1 & 1.32(5)  & 0.15(3) & 1.78(5) & 0 & -0.14(2) \\
SWT + $1/S$ & 1 & 1.35  & 0.15 & 1.59 & 0 & -0.14 \\
\end{tabular}
\end{ruledtabular}
\end{table}

\subsection{Spin-wave theory with $1/S$ corrections for
easy-axis triangular layers}
\label{sec_swt}
In the absence of anisotropy the 2D triangular $J_1-J_2$
antiferromagnet in the collinear phase has been shown to exhibit a
strong renormalization of the dispersion relation due to quantum
fluctuations. Including the effects of fluctuations to order $1/S$
in spin wave theory\cite{Chubukov_1992} generates a gap at the
soft points where the LSWT calculation predicts unphysical gapless
modes, as illustrated in Fig.~\ref{fig_Panneled_Dispersion_J1J2}a) solid line is SWT+$1/S$ and
dashed is LSWT.

In Appendix~\ref{sec_appendixB} we have
calculated the $1/S$ corrections to the dispersion relation for
the triangular antiferromagnet when an easy-axis anisotropy is
also present and the results are plotted in
Fig.~\ref{fig_Panneled_Dispersion_J1J2}b).
The energy at the soft point
is now only marginally renormalized up in energy because the
easy-axis anisotropy generates a gap in the spectrum and this
reduces significantly the zero-point quantum fluctuation effects.
Using the 2D dispersion relation at order $1/S$ and including also
the inter-layer coupling
the best parameter values obtained from comparison to the data are
very similar to those found using LSWT and are shown in Table~\ref{fit_table}.
The dispersion relation at order $1/S$ for these parameters is plotted
in Fig.~\ref{fig_Panneled_Dispersion_J1J2}d) (solid line) and is very
similar to the LSWT dispersion for the uncorrected parameters
(dashed line). The powder-averaged spectrum is also similar to
Figs.~\ref{fig_lswt}a-b) for the uncorrected parameters and is
therefore not reproduced here.

In converging to the above set of parameter values for the
microscopic Hamiltonian we have used two constraints imposed by
the data i) the inter-layer dispersion is relatively small, $\sim
10\%$ of the bandwidth, as inferred from the observed small
dispersion in the low-energy boundary of the magnetic inelastic
scattering, and ii) the spin gap is relatively large, $\sim 30\%$
of the bandwidth, which in turn implies a relatively large
easy-axis anisotropy. Specifically, subtracting the estimated
inter-layer dispersion magnitude from the observed maximum energy
leads us to search for a 2D dispersion shape where the gap energy
divided by the maximum, i.e. the reduced gap $\Delta/\omega_{\rm
max}$, is $\sim 25\%$.

\begin{figure}[t]
\begin{center}
\includegraphics[width=7.8cm,  bbllx=153,bblly=289,bburx=377,
 bbury=444] {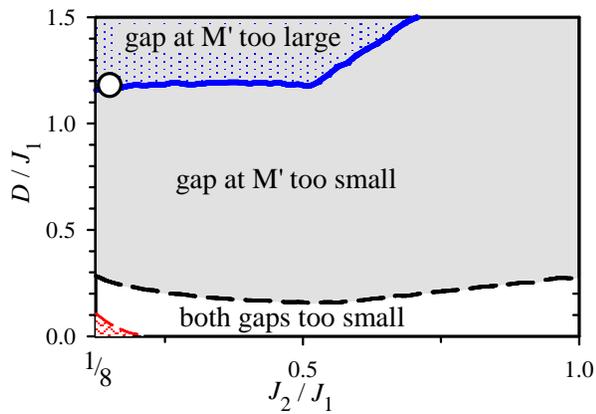}
\caption{\label{fig_reduced_gap} (Color online) Constraints on the
parameters $J_2/J_1$ and ${\cal D}/J_1$ determined from the spin
gap measured in inelastic neutron scattering experiments. Gaps at
M and M$'$ are calculated within SWT+$1/S$ discussed in
Appendix~\ref{sec_appendixB}, and normalized to the maximum energy
for spin excitations. The upper solid line divides solutions where
the calculated gap at M$'$ is larger than that seen in experiment,
from those where it is smaller. The point $J_2/J_1 \approx 0.15$,
${\cal D}/J_1 \approx 1.2$ which gives the best fit overall fit to
experiment is marked with an open circle. The region of small
${\cal D}$ for which neither the gap at M nor M$'$ are large
enough to fit experiment is bounded by a dashed line. Linear spin
wave calculations give similar lines, slightly shifted. [The small
region of parameter space ${\cal D}/J_1 \gtrsim 0$, $J_2/J_1
\gtrsim 1/8$ for which spin wave theory is ill-conditioned is
shown by the red hatched area.]}
\end{center}
\end{figure}

The dispersion of the microscopic Hamiltonian in eq.\
(\ref{eq:Hamiltonian}) has the minimum either at the Bragg peak
position M, or at the soft point M$'$. This places a strong
constraint on model parameters, since both reduced gaps must be at
or above the threshold value of $25\%$, with at least one gap
precisely at the threshold. In Fig.\ \ref{fig_reduced_gap} we map
out these constraints for in ($J_2/J_1$,${\cal D}/J_1$) parameter
space, with gaps calculated using spin wave theory with $1/S$
corrections, as described in Appendix~\ref{sec_appendixB}.
Possible solutions lie on the solid line dividing solutions where
the reduced gap $\Delta_{{\rm M}'}/\omega_{\rm max} > 25$\% from
those where $\Delta_{{\rm M}'}/\omega_{\rm max} < 25$\%.

On this line other features in the data such as the observed slope of the
low-$Q$ dispersion further narrow down possible parameter values,
leading to the best fit values. So within the minimal microscopic
Hamiltonian considered the requirement for quantitative agreement
for the absolute gap relative to the maximum energy requires a
relatively large anisotropy, and in this case the dispersion
minimum is at the soft point M$'$ as illustrated in Fig.\
\ref{fig_Panneled_Dispersion_J1J2}d) (solid line).

\begin{figure}[t]
\begin{center}
\includegraphics[width=7.5cm,  bbllx=61,bblly=228,bburx=476,
 bbury=564] {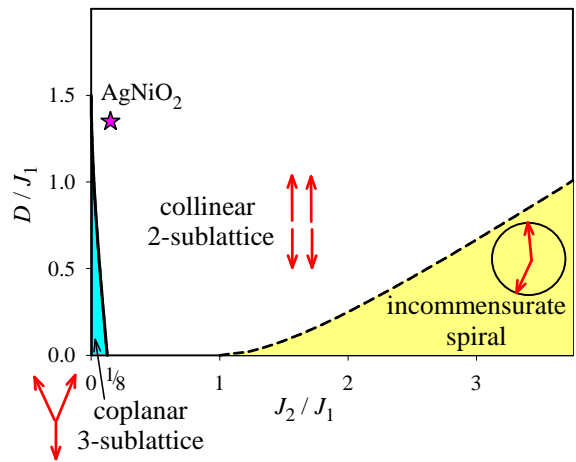}
\caption{\label{fig_J2-D_phase_diagram} (Color online) Ground
state phase diagram of the easy-axis Heisenberg model on a
triangular lattice as a function of $J_2/J_1$ and ${\cal D}/J_1$, calculated from mean field 
theory.  The parameters $J_2/J_1 \approx 0.15$, ${\cal D}/J_1
\approx 1.3$ deduced from linear spin wave theory fits to
inelastic neutron scattering data place AgNiO$_2$ within the
collinear AF phase, not far from the boundary with a coplanar
3-sublattice state. }
\end{center}
\end{figure}

\section{Discussion and conclusions}
\label{sec_Conclusions}
We have reported inelastic neutron scattering measurements of the
powder-averaged spin dynamics in the layered hexagonal
antiferromagnet $2H$-AgNiO$_2$, which has a collinear alternating
stripe ordered ground state. A broad band of magnetic inelastic
scattering is observed above a finite gap indicating strongly
dispersive magnetic excitations. The observed modulations in the
boundaries of the magnetic scattering could be well explained by a
gapped and predominantly two-dimensional dispersion relation,
suggesting magnetically weakly-coupled triangular layers. For a
quantitative analysis we have considered a minimal spin
Hamiltonian in eq.~(\ref{eq:Hamiltonian}) for the localized Ni1
ions ($S=1$) located on an ideal triangular lattice and have
neglected the coupling to the itinerant electrons on the
surrounding Ni sites.  Within this localized model we considered
first- and second-neighbour antiferromagnetic couplings $J_1$ and
$J_2=0.15 J_1$ in the triangular layers, a strong easy axis
anisotropy modelled by a single ion term ${\cal D}\simeq 1.3J_1$
and weak interlayer couplings $J_z'=-0.1 J_1$ and have found that
this minimal model can account quantitatively for all the observed
wavevector-dependence of the boundaries of the powder-averaged
spectrum.

The deduced values of the magnetic interactions provide a natural
explanation for the stability of the observed collinear
alternating stripe magnetic order. The large value for the
easy-axis anisotropy strongly modifies the physics of an
antiferromagnet on a triangular lattice. Firstly, it stabilizes
collinear order at the expense of coplanar states like the
120$^\circ$ spiral ground state of a simplest nearest-neighbour
Heisenberg model. And secondly, it opens a gap to spin excitations
which strongly suppresses quantum zero-point fluctuations. These
are essential for the ``order from disorder'' selection of
collinear order in the {\it isotropic} $J_1$--$J_2$ model
considered by Chubukov and Jolicoeur~\cite{Chubukov_1992}.
But in the easy-axis $J_1$--$J_2$ model relevant to AgNiO$_2$,
fluctuations lead only to small quantitative corrections to the
spin wave dispersion.

We have calculated these corrections explicitly to ${\mathcal
O}(1/S)$ and find that, for most purposes, they can safely be
neglected as they only lead to a very modest renormalization of
the spin wave dispersion --- and the parameters inferred from it
--- as illustrated in Table~\ref{fit_table}. Competing
interactions like $J_2$ also play a secondary role, selecting the
observed form of collinear order from the vast manifold of Ising
ground states on a triangular lattice. The extent of the collinear
AF phase as a function of ($J_2/J_1$,${\cal D}/J_1$) can easily be
estimated within mean field theory. The results of this analysis
are shown in  Fig.~\ref{fig_J2-D_phase_diagram}.   The parameter
set deduced above, with strong anisotropy ${\cal D}$ but
relatively small second neighbour interaction $J_2$ places
AgNiO$_2$ within the collinear AF phase --- as required --- but
not far from the boundary with a coplanar 3-sublattice state.

At present, the origin of this strong easy-axis anisotropy ---
which may be of single-ion or exchange character --- is unclear.
AgNiO$_2$ is a highly anisotropic material, and the crystal fields
at the magnetic sites will reflect this. However, relativistic
(i.e. spin-orbit coupling) effects in Ni are not usually strong,
so the absolute size of anisotropy is surprising. The more massive
Ag ions, which have stronger spin-orbit coupling, are
non-magnetic, and thought to play a negligible role in
interactions between the Ni$^{2+}$ sites.   However, given the
partially itinerant nature of Ni electrons, a simple local-moment
model like eq.~(\ref{eq:Hamiltonian}) is probably unlikely ever to
offer a complete description of the magnetism in AgNiO$_2$.

Further band-structure calculations based on the model developed in
Ref.~\onlinecite{AgNiO2_prl} could potentially test the proposed
minimal Hamiltonian and give further insight into the microscopic
origin of the exchange interactions and easy-axis anisotropy.   Of
particular interest would be the contribution of the itinerant Ni2
and Ni3 sites which surround the strongly-magnetic Ni1 sites to
magnetic excitations, and their role in mediating interactions
between localized Ni1 sites. Even though the dispersion of the
boundaries of the powder-averaged data are well explained by
eq.~(\ref{eq:Hamiltonian}), there are discrepancies in the intensity
distribution as a function of energy and wavevector [see
Fig.~\ref{fig_lswt_fits}E], and these may contain information about
the itinerant electrons.  It is also interesting to ask whether the
weak, but long-range interlayer interaction $J_z^\prime$, could be
of RKKY origin ? This is a promising avenue for future research, and
recent transport experiments on AgNiO$_2$ in high magnetic field
reinforce the idea that charge carriers couple strongly to the
magnetic order.\cite{A_Coldea}

Even without these complications, the minimal microscopic model we
have proposed has a number of very interesting features. In
particular it predicts a spin wave dispersion with global minima
not at magnetic Bragg wavevectors but at symmetry-related soft
points. This prediction could be tested by future inelastic
neutron scattering experiments on single crystals or by ESR
experiments, which probe the excitations at the $\Gamma$ point,
which by periodicity have the same energy as those at the magnetic
Bragg wavevector. We note that the unusual dispersion relation we
propose with global minima at multiple soft points may lead to
non-trivial magnetically-ordered phases in applied magnetic fields
that overcome the spin gap and future studies of the phase diagram
in high applied magnetic field along the easy-axis could show a
rich behaviour that would be fruitful to explore
experimentally\cite{Amalia} and theoretically.\cite{Seabra}

\section{Acknowledgements}
\label{sec_acknowledgements}
We acknowledge useful discussions with A.~V.~Chubukov,
A.~I.~Coldea, K.~Damle, M.~Enderle, L.~Seabra and A.~Vishwanath.
The research was supported in part by EPSRC UK grants GR/R76714/02
(RC), EP/C51078X/2(EW), and EP/C539974/1 (NS), a CASE award from
the EPSRC and ILL (EMW), and the EU programme at the ILL. NS and
PA gratefully acknowledge the support of the guest program of
MPI-PKS Dresden, where part of this work was carried out.

\begin{figure}
\begin{center}
\includegraphics[width=7.9cm,  bbllx=55,bblly=288,bburx=541,
 bbury=554] {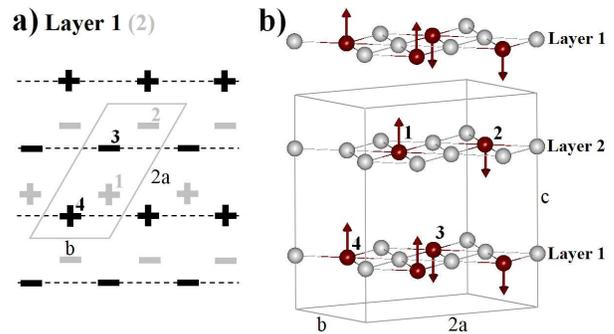}
\caption{\label{fig_Spin_wave_sketch}(Color online) Magnetic
structure of $2H$-AgNiO$_2$. a) In each layer spins form
alternating ferromagnetic stripes (dashed lines), $\pm$ symbols
indicate the projection of the spin moments along the $c$-axis.
Stripes are parallel between adjacent layers (thick black symbols
for layer 1 and faint gray for layer 2) but have an offset because
of the structural arrangement of Ni1 ions. Labels 1-4 refer to the
four magnetic sublattices (two for each layer) in the magnetic
unit cell (solid contour), which is doubled along $a$-axis
compared to the crystallographic unit cell. b) 3D view of the
magnetic structure showing the orientation of the spin moments
(thick arrows) of the Ni1 ions (dark brown spheres) in the two
layers in the unit cell. Light gray spheres are Ni2 and Ni3 ions,
assumed non-magnetic. The box shows the 3D magnetic unit cell.}
\end{center}
\end{figure}

\appendix
\section{Linear spin-wave dispersions for coupled triangular layers}
\label{sec_appendix}

Here we outline the derivation of the linear spin-wave dispersion
relations for a model of stacked antiferromagnetic triangular
layers with collinear stripe order. The experimentally-determined
magnetic structure in $2H$-AgNiO$_2$ is shown in Fig.\
\ref{fig_Spin_wave_sketch}. In each triangular Ni layer the
ordered spins form alternating ferromagnetic rows described by two
magnetic sublattices (up/down), leading to a total of 4 magnetic
sublattices for the two layers in the magnetic unit cell (box in
Fig.~\ref{fig_Spin_wave_sketch}b). After a Holstein-Primakoff
transformation from spin to magnon operators and Fourier
transformation the spin Hamiltonian in eq.~(\ref{eq:Hamiltonian})
becomes
$$
\mathcal{H}=\sum_{\bm q}\mathsf{X}^{\dagger}H\mathsf{X}
+E_0\label{Ham_equ}$$ where $E_0$ is a constant and terms higher
than quadratic are neglected. The sum extends over all wavevectors
${\bm q}$ in the first magnetic Brillouin zone. The ${\bm q}$
dependence of the operator matrix $\mathsf{X}$ and of the
Hamiltonian matrix $H$ is implicit. The operator matrix
$\mathsf{X}^{\dagger}$ is given by
$$\mathsf{X}^\dagger=\left[\alpha^\dagger_{\bm q} ~,~
\gamma^\dagger_q ~,~ \beta_{-{\bm q}} ~,~ \epsilon_{-{\bm
q}}\right],$$ where $\alpha$, $\beta$, $\epsilon$ and $\gamma$
refer to the magnetic sublattices 1, 2, 3, 4 (in this order) and
$\alpha^\dagger_{\bm q}$ creates a plane-wave magnon mode on
sublattice 1, and so on. The Hamiltonian matrix in this operator
basis is
\begin{equation}
\label{eq_App_Ham_matrix} H=\left[\begin{array}{llll}
A&   B&  C & D^* \\
B^*&  A&  D& C \\
C& D^* & A& B \\
D& C & B^* & A \\
\end{array}\right]
\end{equation}
where
\begin{displaymath}
\begin{array}{l l}
    A&=2S \left[ J_1 \cos(2\pi k)+J_1+J_2 \cos(2\pi(2h+k))+J_2 \right. \\
    &\, \left. +J_z' \cos(2\pi l)-J_z'-J_{z}\right]+2S{\cal D}\left(1-\frac{1}{2S}\right)\\
    B&=4SJ_{z} \cos(\pi k)\cos(\pi l)~ \zeta\\
    C&=4SJ_1 \cos(\pi k)\cos(\pi(2h+k))\\
    &\,+4SJ_2\cos(3 \pi k)\cos(\pi (2h+k))\\
    D&=2SJ_{z}\cos(\pi l) ~\zeta^2\\
    \zeta&=e^{-(2h+k) \pi i/3},\\
\end{array}
\end{displaymath}
where $(h,k,l)$ are components of the wavevector ${\bm q}$ in
units of the reciprocal lattice of the hexagonal unit cell shown
in Fig.~\ref{fig_Structure}b). Note that the very last term in
function $A$ above, $2S{\cal D}\times\frac{-1}{2S}$ formally comes
in spin-wave theory as a higher order ($1/S$) correction term, but
we have included it here following common
convention,\cite{kaganov_87} since it originates from a quadratic
term in boson operators.

Diagonalizing the Hamiltonian in eq.~(\ref{eq_App_Ham_matrix})
using standard methods\cite{white} gives two doubly-degenerate
modes with dispersions given by
\begin{equation}
\begin{array}{ll}
(\omega^{\pm}_{\bm q})^2&=A^2+BB^*-C^2-DD^*\pm \\
&{\rm sgn}\left(B/\zeta\right)\sqrt{4|AB-CD^*|^2-|B^*D^*-BD|^2}
\end{array}
\label{eq_App_dispersion}
\end{equation}
where $\pm$ stands for optic and acoustic modes between the two
layers and ${\rm sgn}(x)$ is the sign function. The above
expressions for the dispersion relations also hold by periodicity
for a general wavevector ${\bm Q}=(h,k,l)$ outside the first
magnetic Brillouin zone [the two modes are degenerate when $B=0$
and the prefactor ${\rm sgn}\left(B/\zeta\right)$ in front of the
square root is required to ensure continuity of the two distinct
modes on both sides of a degenerate point].

For decoupled triangular layers with no anisotropy
($J_z=J_z'={\cal D}=0$ the dispersion is plotted in Fig.\
\ref{fig_Panneled_Dispersion_J1J2}a) (dashed line) and for the
general case is plotted in Fig.\
\ref{fig_Panneled_Dispersion_J1J2}c)(dashed and dash-dotted
lines). The finite easy-axis anisotropy ${\cal D}>0$ leads to a
gap in the spectrum. The gap above the magnetic Bragg peaks
increases as a power law with increasing anisotropy $\omega^-_{\rm
M}=2S\sqrt{\tilde{{\cal D}}\left(4J_1+4J_2+2J_z+\tilde{{\cal
D}}\right)}$, where $\tilde{{\cal D}}={\cal
D}\left(1-\frac{1}{2S}\right)$. However, the gap is near-linear in
anisotropy at the soft point M$^{\prime}$, $\omega^{\pm}_{{\rm
M}'}=2S\sqrt{\tilde{\cal D}\left(\tilde{\cal D}-2J_z\right)}$, so
in the limit of weak couplings between the layers the minimum gap
is always at the soft point M$'$ and not at the Bragg wavevector M
[see Fig.\ \ref{fig_Panneled_Dispersion_J1J2}b)].

The dynamical correlations (per Ni1 spin) are obtained as
\begin{eqnarray}
S^{xx}&({\bm Q},\omega) =\mathcal{C} \frac{S}{8}\left[
\frac{\left|W(-\omega_+)+X(-\omega_+)+Y(-\omega_+)+Z(-\omega_+)\right|^2}{N(-\omega_{+})}+\right.\nonumber  \\
& \left. \frac{\left|W( \omega_+)+X( \omega_+)+Y(
\omega_+)+Z(\omega_+)\right|^2}{N( \omega_{+})}\right]
~G(\omega-\omega_+)\nonumber \\
&+\mathcal{C}\frac{S}{8}\left[
\frac{\left|W(-\omega_-)+X(-\omega_-)+Y(-\omega_-)+Z(-\omega_-)\right|^2}{N(-\omega_{-})}+\right.\nonumber \\
& \left.
\frac{\left|W(\omega_-)+X(\omega_-)+Y(\omega_-)+Z(\omega_-)\right|^2}{N(\omega_-)}\right]
~G(\omega-\omega_-) \label{eq_App_Sqw}
\end{eqnarray}
where $G(\omega-\omega_{\pm})$ is a Gaussian with a finite width
to model the instrumental resolution, we use the shorthand
notation $\omega_{\pm}=\omega^{\pm}_{\bm Q}$ and the functions
$W$, $X$, $Y$, $Z$ and $N$ are given by
\begin{displaymath}
\begin{array}{ll}
W(\omega)&=-(A+\omega)(A^2+BB^*-C^2-DD^*-\omega^2)+\\
         &\quad 2ABB^*-C(B^*D^*+BD)
\end{array}
\end{displaymath}
\begin{displaymath}
\begin{array}{ll}
X(\omega)&=C(A^2+BB^*-C^2+DD^*-\omega^2)-\\
         &\quad A(B^*D^*+BD)-\omega(B^*D^*-BD)
\end{array}
\end{displaymath}
\begin{displaymath}
\begin{array}{ll}
Y(\omega)&=B^*\left[(A+\omega)^2-BB^*+C^2\right]-\\
         &\quad 2C(A+\omega)D+BD^2
\end{array}
\end{displaymath}
\begin{displaymath}
\begin{array}{ll}
Z(\omega)&=D(A^2+C^2-DD^*-\omega^2)+B^{*2}D^*-\\
         &\quad 2AB^*C
\end{array}
\end{displaymath}
\begin{displaymath}
\begin{array}{ll}
N(\omega)&=|-WW^*+XX^*-YY^*+ZZ^*|.
\end{array}
\end{displaymath}
The overall intensity prefactor $\mathcal{C}$ in eq.\
(\ref{eq_App_Sqw}) is 1 in LSWT and is considered here as a
variable parameter in the comparison with the experimental data in
order to account (in a first approximation) for a possible overall
intensity renormalization compared to LSWT. We note that the above
expressions simplify considerably in the absence of the interlayer
coupling $J_z$, in that case the two magnon modes in eq.\
(\ref{eq_App_dispersion}) become degenerate with dispersion
$\omega_{\bm{Q}}=\sqrt{A^2-C^2}$ and the dynamical correlations in
eq.\ (\ref{eq_App_Sqw}) become
\begin{equation} S^{xx}(\bm{Q},\omega)={\cal C} \frac{S}{2}
\frac{A-C}{\omega} G(\omega-\omega_{\bm{Q}}).
\end{equation}

For completeness we also quote the obtained transformation matrix
${\cal S}$ to the normal operator basis $\mathsf{X}'=\mathcal{
S}^{-1}\mathsf{X}$ in which the Hamiltonian is diagonal
$$\mathcal{S}=\left[
\begin{tabular}{cccc}
$\overline{W}(\omega_-)$ & $\overline{W}(\omega_+)$ & $\overline{W}(-\omega_-)$ & $\overline{W}(-\omega_+)$\\
$\overline{Y}(\omega_-)$ & $\overline{Y}(\omega_+)$ & $\overline{Y}(-\omega_-)$ & $\overline{Y}(-\omega_+)$\\
$\overline{X}(\omega_-)$ & $\overline{X}(\omega_+)$ & $\overline{X}(-\omega_-)$ & $\overline{X}(-\omega_+)$\\
$\overline{Z}(\omega_-)$ & $\overline{Z}(\omega_+)$ & $\overline{Z}(-\omega_-)$ & $\overline{Z}(-\omega_+)$\\
\end{tabular}
\right]$$ where we use the shorthand notation
$\overline{W}(\omega)=W(\omega)/\sqrt{N(\omega)}$ and so on. This
matrix satisfies the eigenvalue equation
$\mathsf{g}H\mathcal{S}=\mathcal{S}\mathsf{g}H'$ and the
normalization condition
$\mathcal{S}\mathsf{g}\mathcal{S}^{\dag}=\mathsf{g}$, where
$\mathsf{g}$ is the commutator matrix for the operator basis
defined by
$$\mathsf{g}=\mathsf{X}(\mathsf{X}^*)^{T}-(\mathsf{X}^{*}\mathsf{X}^{T})^{T}=
 \left[
 \begin{array}{cccc}
 1&0&0&0\\
 0&1&0&0\\
 0&0&-1&0\\
 0&0&0&-1\\
 \end{array}
 \right]$$ and
$$H'= \left[
\begin{tabular}{cccc}
$\omega_-$ & 0 & 0 & 0\\
0 & $\omega_+$  & 0 & 0\\
0 & 0 & $\omega_-$ & 0\\
0 & 0 & 0 & $\omega_+$\\
\end{tabular}
\right]$$ is the Hamiltonian representation in the normal operator
basis $\mathsf{X}'$.

The one-magnon neutron scattering intensity including the
polarization factor and the magnetic form factor is
\begin{equation} S(\bm{Q},\omega)=\left(\gamma r_{\rm
o}\right)^2\left(1+\frac{Q_z^2}{Q^2}\right)\left(\frac{g}{2}
f(Q)\right)^2 S^{xx}(\bm Q,\omega) \label{eq_App_cross_section}
\end{equation}
where $\left(\gamma r_{\rm o}\right)^2$=290.6 mbarns/sr is a
conversion factor to bring the intensity into absolute units of
mbarns/meV/sr, $f(Q)$ is the magnetic form factor for Ni$^{2+}$
ions and we take the $g$-factor $g=2$. $Q_z$ is the component of
the wavevector transfer ${\bm Q}$ along the $c$-axis. A numerical
average of eq.~(\ref{eq_App_cross_section}) over a spherical
distribution of orientations of the wavevector transfer ${\bm Q}$
was performed in order to obtain the orientational-averaged
intensity $S(Q,\omega)$ to be compared with the measured powder
data.

\section{Spin-wave theory with $1/S$ quantum corrections for the
easy-axis $J_1-J_2$ triangular antiferromagnet}
\label{sec_appendixB}
Here we outline calculations of corrections to spin wave
dispersion arising from interactions between magnons at ${\mathcal
O}(1/S)$. Calculations to this order have previously been carried
out by Chubukov and Jolicoeur for the {\it isotropic} $J_1$-$J_2$
Heisenberg model on a triangular lattice~\cite{Chubukov_1992}.
Here we extend this work to the case of a finite easy-axis
anisotropy ${\cal D}$.  To make contact with this earlier work we
use here an orthogonal coordinate system with $x$ along the
magnetic stripes and $y$ transverse to stripes in plane as in
Fig.~\ref{fig_BZ}b) and label the 2D wavevectors as
($k_x$,$k_y$)=($Q_x a$,$Q_y a$) where $Q_{x,y}$ are orthogonal
components (in \AA $^{-1}$) and $a$ is the hexagonal lattice
parameter.

In the usual linear, semi-classical approximation (i.e. neglecting
terms which contain more than two Bose operators) we obtain a
dispersion
\begin{eqnarray}
\omega_{\bm k}=\sqrt{A_{\bm k}^2-B_{\bm k}^2}    \nonumber
\end{eqnarray}
with coefficients
\begin{eqnarray}
A_{\bm k}&=&2SJ_1\left( 1+\cos{k_x} +\alpha+
\alpha\cos\sqrt{3}k_y \right)  + 2S \tilde{\cal D} \nonumber\\
B_{\bm k}&=&4SJ_1 \cos\frac{\sqrt{3}k_y}{2}
\left(\cos\frac{k_x}{2}+\alpha\cos\frac{3k_x}{2}\right)
\label{eq_App_AB}
\end{eqnarray}
where $\alpha = J_2/J_1$ and $\tilde{\cal D} = {\cal D}(1 -
1/(2S))$. We note that our units differ by an overall factor of
two from those in Ref.~\onlinecite{Chubukov_1992}, and that
single-ion easy-axis anisotropy contributes terms to the
linearized Hamiltonian which are formally of order ${\mathcal
O}(1/S)$ relative to usual LSWT.

The leading effect of interactions between spin waves, treated at
a one-loop (i.e. Harteee-Fock) level, is to renormalize the
coefficients of this dispersion to give
\begin{eqnarray}
\epsilon_{\bm k}=\sqrt{(A_{\bm k}+\delta A_{\bm
k})^2-(B_{\bm k}+\delta B_{\bm k})^2}  \nonumber
\end{eqnarray}
where
\begin{eqnarray}
\label{eq_App_deltaAB}
\delta A_{\bm k}&=&\frac{2J_1}{N}\sum_{\bm
p}\left[\frac{\bar{A}_{\bm p}-\bar{\omega}_{\bm
p}}{2\bar{\omega}_{\bm p}}~F(\bm{k},\bm{p})
+ \frac{\bar{B}^2_{\bm p}}{2\bar{\omega}_{\bm p}}\right] \nonumber\\
\delta B_{\bm k}&=&-\frac{2J_1}{N}\sum_{\bm
p}\left[\frac{\bar{A}_{\bm p}-\bar{\omega}_{\bm
p}}{2\bar{\omega}_{\bm p}}~\bar{B}_{\bm k} -\frac{\bar{B}_{\bm
p}}{2\bar{\omega}_{\bm p}}~G(\bm{k},\bm{p})\right]
\end{eqnarray}
Here, $N$ is the total number of spins in the triangular plane,
and the sum on internal momenta $\bm{p}$ extends over the entire
nuclear Brillouin zone (large thick line hexagon in
Fig.~\ref{fig_BZ}a). Following Ref.~\onlinecite{Chubukov_1992}, we
write $\bar{Z}=Z/(2SJ_1)$ ($Z=A,B,\omega$) and introduce the
functions
\begin{eqnarray}
F(\bm{k},\bm{p})&=&(1-\cos{k_x})(1-\cos{p_x}) -2(1+\alpha) \nonumber\\
& &+\alpha(1-\cos{\sqrt{3}k_y})(1-\cos{\sqrt{3}p_y})
-\frac{2{\cal D}}{J_1}   \nonumber\\
G(\bm{k},\bm{p})&=&\cos\frac{\sqrt{3}k_y+k_x}{2}
\cos\frac{\sqrt{3}p_y+p_x}{2} \nonumber\\
& &+ \cos\frac{\sqrt{3}k_y-k_x}{2}
\cos\frac{\sqrt{3}p_y-p_x}{2} \nonumber\\
& &+ \alpha\left[ \cos\frac{\sqrt{3}k_y+3k_x}{2}
\cos\frac{\sqrt{3}p_y+3p_x}{2} \right. \nonumber\\
& &+ \left. \cos\frac{\sqrt{3}k_y-3k_x}{2}
\cos\frac{\sqrt{3}p_y-3p_x}{2} \right]
\label{eq_App_FG}
\end{eqnarray}
Anisotropy ${\cal D}$ enters $\delta A_{\bm k}$ explicitly
through $F(\bm{k},\bm{p})$, 
but its main effect is to introduce a gap into the non-interacting
dispersion $\bar{\omega}_{\bm k}$, which acts as an infra-red
cutoff in {\it all} denominators.  In order not to mix different
orders in $1/S$, we set $\tilde{\cal D} \to {\cal D}$ in
calculations of $\delta A_{\bm k}$ and $\delta B_{\bm k}$.

To obtain quantitative results for comparison with experiment, the
sums in eq.~(\ref{eq_App_deltaAB}) were evaluated numerically. The
resulting corrections to the linear spin-wave dispersion are
illustrated first for the case of no anisotropy in
Fig.~\ref{fig_Panneled_Dispersion_J1J2}a) (solid line). In this
case we exactly reproduce the results of
Ref.~\onlinecite{Chubukov_1992} : LSWT (dashed line) predicts a
gapless mode at the soft point M$'$ (as an indirect consequence of
the macroscopic classical ground state degeneracy), however
quantum corrections lift the classical degeneracy by stabilizing
the collinear stripe order and generating a spontaneous gap at the
soft points while maintaining a gapless Goldstone mode at the
magnetic Bragg wavevectors.

In the presence of anisotropy the situation is somewhat different.
Firstly the collinear order observed in AgNiO$_2$ is now stable at
a {\it classical} level for a wide range of ${\cal D}$~and~$J_2$
(c.f. Fig.~\ref{fig_J2-D_phase_diagram}). And secondly, {\it all}
spin excitations are now gapped, even in the absence of
fluctuation effects [see
Fig.~\ref{fig_Panneled_Dispersion_J1J2}b)]. Expanding about
collinear order, at a semiclassical level, the gap at the Bragg
wavevectors M $= (0, 2\pi/\sqrt{3}) $ grows as
$$
\Delta_{\rm M} \approx 2S\sqrt{\tilde{\cal
D}(4J_1+4J_2+\tilde{\cal D})}
$$
whereas at the soft points ${\rm M}^\prime= (\pi, \pi/\sqrt{3})$
it increases only linearly as $2S\tilde{\cal D}$.   As a result,
for finite ${\cal D}$, the minimum in the spin wave dispersion
occurs on the zone boundary at the soft point ${\rm M}^\prime$,
and {\it not} at the magnetic ordering vector ${\rm M}$.  Once
$1/S$ corrections are included, a gap is generated dynamically at
M$'$ for ${\cal D}=0$, and there is a threshold anisotropy value
${\cal D}_c$ above which the global minimum gap moves from ${\rm
M}$ to the soft point  ${\rm M}'$.

For parameters relevant to $2H$-AgNiO$_2$ (see
Table~\ref{fit_table}) the gap to spin excitations is large
throughout the BZ, with a minimum at M$'$ (${\cal D}_c \approx
0.21$~meV for this parameter set). Quantum fluctuations lead only
to small quantitative corrections to the dispersion --- compare
the dashed to the solid line in
Fig.~\ref{fig_Panneled_Dispersion_J1J2}b) --- and fits to LSWT are
therefore justified {\it a posteriori}. A further indication that
quantum fluctuations are small is the fact that the zero-point
reduction of the sublattice magnetization
\begin{eqnarray}
\delta m_{\sf S} = \frac{2}{N}\sum_{\bm p}\frac{A_{\bm p}-\omega_{\bm p}}{2\omega_{\bm p}}
\end{eqnarray}
is calculated to be small  : $\delta m_{\sf S}=0.038$ for the parameters
given in Table~\ref{fit_table}.

We note that this is {\it not} true in the limit of small
anisotropy ${\cal D} \to 0$ and small NNN couplings \mbox{$J_2/J_1
\to 1/8$}, in which case the system is close to a quantum critical
point. In this case zero-point fluctuations are very strong and
the correction to the sublattice magnetization diverges within
LSWT.    A consequence of this is that $1/S$ corrections drive the
gap
\begin{eqnarray}
\Delta_\Gamma &\approx& 4\sqrt{(J_1S + J_2S) {\cal D} \left[S -
1/2 - \delta m_{\sf S} \right]}
\end{eqnarray}
at the $\Gamma$ point ${\bf k} = (0,0)$ {\it imaginary} for a
small but finite region of parameters ${\cal D}/J_1 \gtrsim 0$,
\mbox{$J_2/J_1 > 1/8$}  --- c.f. Fig.~\ref{fig_reduced_gap}.
Accurate treatment of this small ${\cal D}$ regime would require a
more sophisticated self-consistent spin wave theory, which is not
warranted for AgNiO$_2$.

For the purposes of comparing the spin-wave calculations at
${\mathcal O}(1/S)$ with the {\it envelope} of scattering in $(|{\bm
Q}|, \omega)$ it is sufficient to replace $A_{\bm k}$, $B_{\bm k}$
and $\omega_{\bm k}$ by their corrected values in the expression
for the scattering intensity
\begin{eqnarray}
S^{xx}(\bm{Q},\omega)
  &=& {\cal C} \frac{S}{2} \frac{A_{\bm k}-B_{\bm k}}{\omega_{\bm k}}~\delta(\omega-\omega_{\bm
  k}).
\end{eqnarray}

Finally, the interlayer coupling $J_z'$ can easily be included in
these calculations by modifying Eq.~(\ref{eq_App_AB}) as follows
\begin{eqnarray}
A_{\bm k} &\to&  A_{\bm k}  + 2SJ_{z}'\left(\cos k_z-1\right),
\nonumber
\end{eqnarray}
Here $k_z=Q_z c$ where $Q_z$ is the c-axis wavevector component in
\AA$^{-1}$. At order ${\cal O}(1/S)$ the interlayer couplings
modify Eq.~(\ref{eq_App_FG}) to read
\begin{eqnarray}
F(\bm{k},\bm{p}) & \to & F(\bm{k},\bm{p}) + \frac{J_z'}{J_1}
(1-\cos{2k_z})(1-\cos{2p_z}), \nonumber
\end{eqnarray}
however since this change is relatively very small for simplicity
we neglect it in the analysis and only consider the explicit
effects of $J_z'$ through changes in the dispersion relation at
linear order in SWT. The resulting three-dimensional spin wave
dispersion and corrections are shown in
Fig.~\ref{fig_Panneled_Dispersion_J1J2}d).


\end{document}